\documentclass[a4paper,11pt]{article}
\pdfoutput=1 

\usepackage{jcappub} 

\usepackage[T1]{fontenc} 

\usepackage{booktabs}
\usepackage{float}
\title{\boldmath The H.E.S.S. Gravitational Wave Rapid Follow-up Program}

\author[a,*]{Halim Ashkar}
\author[a]{, Francois Brun}
\author[b]{, Matthias F\"u{\ss}ling}
\author[c]{, Clemens Hoischen}
\author[b]{, Stefan Ohm}
\author[b]{, Heike Prokoph}
\author[a,d]{, Patrick Reichherzer}
\author[a,*]{, Fabian Sch\"ussler}
\author[a,e,*]{and Monica Seglar-Arroyo}
\note[*]{Corresponding author.}

\affiliation[a]{IRFU, CEA, Université Paris-Saclay, F-91191 Gif-sur-Yvette, France}
\affiliation[b]{DESY, D-15738 Zeuthen, Germany}
\affiliation[c]{Institut f\"ur Physik und Astronomie, Universit\"at Potsdam, Karl-Liebknecht-Strasse 24/25, D 14476 Potsdam, Germany}
\affiliation[d]{Plasma-Astroparticle Physics, Faculty for Physics \& Astronomy, Ruhr-Universit\"at Bochum, 44780 Bochum, Germany}
\affiliation[e]{Laboratoire d’Annecy de Physique des Particules, Univ.Grenoble Alpes, Université Savoie Mont-Blanc, CNRS/IN2P3, F-74941 Annecy, France}

\emailAdd{halim.ashkar@cea.fr}
\emailAdd{monica.seglar-arroyo@lapp.in2p3.fr}
\emailAdd{fabian.schussler@cea.fr}

\abstract{Gravitational Wave (GW) events are physical processes that significantly perturbate space-time, e.g. compact binary coalescenses,  causing the production of GWs. The detection of GWs by a worldwide network of advanced interferometers offer unique opportunities for multi-messenger searches and electromagnetic counterpart associations. While carrying extremely useful information, searches for associated electromagnetic emission are challenging due to large sky localisation uncertainties provided by the current GW observatories LIGO and Virgo. Here we present the methods and procedures used within the High Energy Stereoscopic System (H.E.S.S.) in searches for very-high-energy (VHE) gamma-ray emission associated to the emission of GWs from extreme events. To do so we create several algorithms dedicated to schedule GW follow-up observations by creating optimized pointing paterns. We describe algorithms using 2-dimensional GW localisation information and algorithms correlating the galaxy distribution in the local universe, by using galaxy catalogs, with the 3-dimensional GW localisation information and evaluate their performances. The H.E.S.S. automatic GW follow-up chain, described in this paper, is optimized to initiate GW follow-up observations within less than 1 minute after the alert reception. These developements allowed H.E.S.S. observations of 6 GW events out of the 67 non-retracted GW events detected during the first three observation runs of LIGO and Virgo reaching VHE $\gamma$-ray coverages of up to 70\% of the GW localisation.}

\begin{document}
\maketitle
\flushbottom

\section{Introduction}

In 2015, the first detection of Gravitational Waves (GWs) emanating from the inward spiral of two stellar-mass black holes, GW150914~\citep{GW150914}, opened the door to a new era in multi-messenger astrophysics by adding the direct detection of GWs to the list of astrophysical messengers. Nearly two years later, GW170817, a GW event resulting from the merger of a binary neutron star (BNS), was associated to a short $\gamma$-ray burst (GRB) \citep{Abbott_2017}. This event triggered the most extensive multi-wavelength observation campaign in history, leading to the detection of associated emission ranging from the radio up to the X-ray band. Amongst its plentiful implications, GW170817 has established the link between the emission of GWs and short GRBs through BNS mergers. \citep{VHE-BNS} predicts the possibility of long-lasting powerful high energy emission from neutron star mergers involving a short GRB detectable by current IACTs. These emissions are expected to be longer but weaker for off-axis observers. \citep{sGRB_GeV_TeV} argues that short GRBs can reach synchrotron self-Compton powered GeV to TeV afterglow flux that can be detected by current and future TeV instruments in certain scenarios~\citep{sGRB_GeV_TeV}. Other short GRBs have been observed in high energies~\citep{Fermi_GRB} especially GRB 090510 that was deteced by Fermi up to several 10s of GeVs~\citep{Fermi_GRB090510} and GRB 160821B for which MAGIC found evidence of $\gamma$-ray emission above 0.5 TeV~\citep{magic_GRB160821B}. Another milestone in our understanding of GRBs was reached with the detection by the High Energy Stereoscopic System (H.E.S.S.) of very-high-energy (VHE) emission from GRB180702B
~\citep{GRB180720B} and GRB190829A~\citep{GRB190829A_HESS} as well as the detection of GRB190114C~\citep{GRB190114C} by MAGIC. These observations show that the cataclysmic events causing GRBs can trigger efficient particle acceleration, possibly via relavistic jets. The accelerated particles are able to emit $\gamma$-rays in the VHE domain over an extended period of time from the very early (e.g. GRB190114C) to the late afterglow phase (e.g. GRB180720B and GRB190829A). There also seem to be striking similarities between the fading X-ray and VHE $\gamma$-ray flux levels. Except for typically fainter flux level, the X-ray behaviour of short GRBs is similar to long GRBs~\citep{sGRB-lGRB}, thus spawning hopes for possible VHE emission of short GRBs. But, assuming the tentative link between the X-ray and VHE fluxes holds for short GRBs, the fainter fluxes make the afterglow detection of short GRBs in the VHE band even more challenging than for long GRBs. Together with the fact that the emission in the afterglow phase is showing a steady decrease, the need for high sensitivity and rapid follow-up observations in order to detect electromagnetic (EM) counterpart of GWs at VHE energies in the early, bright phase becomes clear. We here rely on this assumption and focus on GW follow-up strategies allowing for fast coverage of GW events directly after a GW detection. Taking into account additional effects that may delay the VHE emission, e.g. off-axis viewing angles as seen in GW170817, multi-wavelength information and dedicated long-term observations ranging from several hours to weeks are necessary. An example of such a campaign is presented in~\cite{EM170817_HESS}.\\

Here we present the methods and procedures used within H.E.S.S. for rapid searches for VHE $\gamma$-ray emission associated to GW events. In Sec.~\ref{sec:GWfollowup}, an overview of the communication between the different observatories involved  and a description of the main information included in the alerts announcing GW detections is provided. We present in  Sec.~\ref{sec:algorithms} the GW follow-up scheduling algorithms developed in this work and used within H.E.S.S. and we discuss their performances, advantages and limitations. In Sec.~\ref{sec:program}, the construction of an automatic exhaustive response scheme to GW alerts based on these algorithms is discussed. We highlight the available choices and describe each step in the decision chain, from the reception of the GW alert to the start of the observation. We illustrate the program with the observations conducted over the last years during the LIGO/Virgo science runs O2 and O3 in Sec.~\ref{sec:osbervations}. The paper concludes with a brief discussion on further improvements and extensions of the H.E.S.S. GW follow-up program that are in preparation for the upcoming science run O4.
\section{Follow-up of gravitational wave alerts with ground-based gamma-ray observatories}
\label{sec:GWfollowup}

The detection of GWs has been achieved so far by the two LIGO interferometers in Livingston and Hanford, in the United States, and the Virgo interferometer, in Italy. The LIGO-Virgo Scientific Collaboration (LVC) has put significant efforts on the real-time analysis of GW signals, their classification as well as the reconstruction of the arrival direction~\citep{singer2016rapid}. An alert is issued and sent to the astrophysics community whenever a significant GW signal is detected, allowing rapid searches for counterparts, i.e. associated electromagnetic emission or neutrinos. During the first (O1, September, 2015 to January, 2016) and the second science run (O2, November, 2016 to August, 2017), alerts were distributed privately. The third science run O3 started on April 1, 2019 and finished on March 27, 2020, with a commissioning break of one month in October 2019. Alerts were issued by LVC through the NASA Gamma-ray burst Coordinates Network (GCN~\citep{GCN}). For each detected signal, several types of GCN notices were emitted at different timescales. The first \textit{preliminary} notices, emitted automatically by the LVC real-time data analysis system during O3, have been made available within 1 to 10 minutes after the arrival of the GW signal. Within the next few hours, either an \textit{initial} or \textit{retraction} notice and a GCN circular were issued. These and all subsequent messages (\textit{updates}), containing refinements on the analysis and localization reconstruction, are human-vetted. Details can be found in~\citep{OpenLVEM}.

The content of the distributed alert messages include a first classification into Compact Binary Coalescence (CBC) or Burst alert, depending on the detection pipeline, the detection time, the GW localisation map, and an event classification into categories of the initial system: binary black hole (BBH), binary neutron star (BNS), binaries comprising a neutron star and a black hole (NSBH) or signal due to terrestrial noise. This information is used to classify and filter these events for potential follow-up observations. The GW localisation map is provided in a HEALPix format~\citep{healpy} that contains four layers of information for each pixel. HEALPix is the acronym for Hierarchical Equal Area isoLatitude Pixelization of a sphere, so the pixelization scheme subdivides the spherical surface in equal area pixels. The resolution of the map is defined by the \textit{N$_{side}$} parameter and the total number of pixels of the map is $N_{pix}=12\times N^2_{side}$. The first layer contains the probability of the GW emission coming from a certain sky direction, in 2D, and the remaining three layers contain event distance information that can be used to obtain a 3D posterior probability~\citep{singer2016going}. The maps distributed in the \textit{preliminary} and \textit{initial} notices are usually computed with the BAYESTAR~\citep{BAYESTAR} algorithm which is used for low latency alerts, and the maps distributed in the \textit{update} notices are computed with LALInference~\citep{LALInference} which provides a more acurate but slower approach. Several techniques have been developed which aim to use this information to guide the search for counterparts to GW events in an efficient way, in particular in the case of telescopes whose field-of-view (FoV) is smaller than the typical GW localisation uncertainty. \\

We here describe the follow-up of GW alerts with Imaging Air Cherenkov Telescopes (IACTs) and especially H.E.S.S. IACTs are ground-based telescopes typically arranged in arrays. Their location on Earth determines the sky visibility. The detection technique in IACTs and the stable operation of the sensitive photo-multiplier based cameras require strong limits on the maximal allowed light levels during observations. All IACTs are therefore operated observatories with low levels of light pollution. These limits also typically restrict observations to astronomical nights without high levels of moonlight falling into the cameras. In order to increase the available observation time and thus the duty cycle of the instruments, all current IACTs have now adopted modes of observation under moderate Moon conditions.  In order to schedule observations under these conditions, Moon and Sun altitude, Moon phase and Moon-to-source distance need to be monitored and considered as additional parameters in the scheduling algorithms.

The energy threshold of observations with IACTs range from a few tens of GeV to roughly 100\,TeV, depending on the zenith angle under which the source is observed. This is due to the absorption of shower light during its passage through the atmosphere, which is higher for larger zenith angles. The effect can be as strong as one order of magnitude of difference in energy threshold when passing to observations at zenith ($\theta{z}$ = 0$^{\circ}$) to observations at large zenith angles ($\theta{z} >$ 60$^\circ$), following an exponential increase. Based on the soft spectrum of GRBs observed~\citep{ackermann2013first}, observations reaching a low energy threshold, i.e. at low zenith angles, are preferred. 

Although the typical FoVs of IACTs reach several degree in radius and are thus sizable compared to observatories operating at other wavelengths, the reconstructed GW source localisation provided by the GW detectors usually exceeds them. Follow-up observations therefore require dedicated scheduling algorithms able to cover the uncertainty regions released by the LVC efficiently and rapidly. In the following we focus on the procedures put in place within H.E.S.S..

H.E.S.S, located at 1800 meters a.s.l. in the Khomas region of Namibia, is a stereoscopic system of five IACTs. The telescope array, sensitive to a broad range of $\gamma$-ray energies from 30 GeV to about 100 TeV, consists of four 12m telescopes (CT1-4) arranged in a square of 120m side length, whose FoV is defined by a circle of about 2.5 degrees radius. A fifth, 28m telescope (CT5), with a FoV defined by a circle of about 1.6 degrees radius, is located in the center of the array. 

Covering the reception and processing of transients alerts, changing the observation schedule and providing preliminary analysis results in real-time, the H.E.S.S. Transients follow-up system~\citep{hessTOOsystem} plays a central role in the execution of any observation program related to transients such as GRBs and GWs. In order to deal with the ever increasing number and variety of information on transient phenomena, the H.E.S.S. Transient follow-up system is fully automatized and does not require human intervention at any stage. The system accepts transients alerts in the VoEvent2.0 format~\citep{allan2017voevent}, which is commonly supplied by current transients brokers. The ToO Alert system subscribes to the NASA Gamma-ray Burst Coordinates Network (GCN~\footnote{\url{https://gcn.gsfc.nasa.gov/}}) alert broker which publishes (among others) alerts from the Fermi-GBM and -LAT, SWIFT-BAT, neutrino detections by IceCube and ANTARES, as well as the alerts from the GW observatories. Further alerts are received via the 4PiSky system~\cite{2016arXiv160603735S}, providing alerts from ASAS-SN, as well as IceCube alerts directly submitted to H.E.S.S..

\begin{figure}
  \centering
\includegraphics[width=0.85\textwidth]{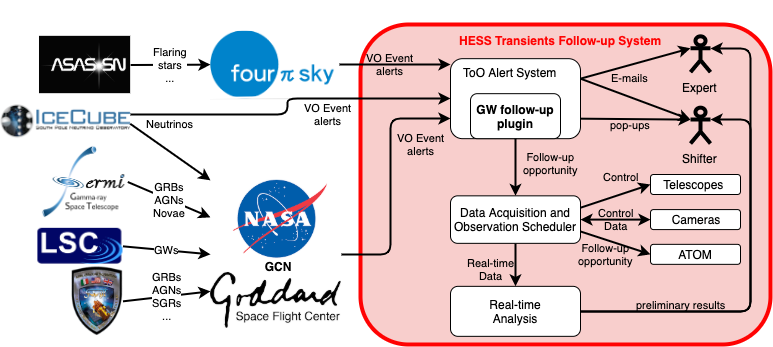}
\caption{Schematic view of the H.E.S.S. ToO alert system providing automatic and real-time reactions to multi-wavelength and multi-messenger alerts.}
\label{fig:VoAlerter}
\end{figure} 

The ToO alert system computes the visibility conditions for incoming alerts and applies selection configurable criteria in order to decide if follow-up observations should take place. In case of a positive selection, the shift-crew and experts are informed via pop-ups on the main screens in the H.E.S.S. control room and via e-mails. If a prompt reaction is possible, the Data Acquisition system of H.E.S.S. will initiate automatic re-positioning of the telescopes by altering the planned schedule. In addition to the H.E.S.S. telescopes, ATOM~\citep{hauser2004atom}, an optical telescope operated at the H.E.S.S. site, is notified in order to provide contemporaneous data in several optical bands. With the onset of the H.E.S.S. observations a real-time analysis of the incoming datastream is started. To provide best sensitivity at low energies where VHE signals of transients are generally expected to be stronger, the real-time analysis is running in monoscopic-mode, i.e. using only data from the large 28m telescope. At predefined signal thresholds, alerts are generated to alert the shift-crew of a possible detection. The results are archived for inspection over the course of the following days. This allows the on-site shift-crew as well as off-site experts to make an informed decision if a ToO observation should be continued or not. An overview of the system is given in Fig.~\ref{fig:VoAlerter} and more details will be provided in an upcoming publication~\citep{hessTOOsystem}.

As the derivation of an optimized scanning pattern to cover the localization of the GWs is a major task, the GW follow-up plugin presents a component of its own within the Transients follow-up system. Similarly, a potential counterpart might be found anywhere in the FoV, posing additional challenges for the real-time analysis. Furthermore, as various types of GW alerts are provided by GW observatories, an adapted processing strategy for each alert type is needed. 

The GW follow-up scheduling algorithms and procedures used within H.E.S.S. for this purpose have changed significantly over the last years during the LVC science runs O1-O3. In preparation of the first science run of the Advanced LIGO interferometers, which started in September 2015, the scheduling foresaw a ring-shape pointing pattern in case of well localized GW events that would surround the most promising region of the localisation uncertainty region provided by the GW detectors. Following the introduction of 3-dimensional localisation uncertainties by the GW instruments in O2~\citep{singer2016going}, the procedure for follow-up observations with H.E.S.S. could be improved further. H.E.S.S. can now rely on a variety of algorithms and a relatively complex decision tree taking into account various parameters of the individual GW event to derive an optimal pointing strategy in the search for associated VHE $\gamma$-ray emission. The algorithms will be explained in the next section.  \\
\section{GW follow-up algorithms}
\label{sec:algorithms}


In general, follow-up strategies are derived with the aim of covering the coordinates from which the GW signal was most probably emitted, and thus the associated multi-wavelength or multi-messenger counterpart emission, as fast as possible ~\citep{abadie2012implementation} and reaching deep observations with a low energy threshold.

  We here present different \textit{tiling strategies}~\citep{2D_tiling} to define a sequential order of individual observations or \textit{pointings}.  The strategy followed in the observation scheduling process proposed in this work falls in the category of {\it greedy scheduling}, i.e. the most \textit{promising} pointing is scheduled at the earliest possible time. The ranking of the observations is based on the probability provided in the GW localization maps, from the highest to the lowest. This natural approach enhances the likelihood of covering the EM counterpart in a shorter period of time, which is assumed to be crucial to detect the multi-wavelength and multi-messenger emission of the remnant.
  

All algorithms use as input the GW localisation map, the alert reception time, the additional parameters characterising the GW event and allow to use flexible telescope configurations through high-level parameters such as the FoV, the maximum allowed zenith angle, the minimum observation run duration and the location of the telescope system. These algorithms include a binned, grid-like scheme to define the individual pointings as well as an unbinned approach using the full 3D localisation information provided by the GW detectors. The output of the developed framework is a detailed observation schedule for a given observational period that can stretch over several nights and which makes optimal use of the available observation time. This observation scheduling information is provided through summary files and figures (see Tab.~\ref{tab:GW170817_schedule} and Fig.~\ref{fig:GW170817_graphics}).\\

All implemented algorithms follow the same general procedure:
\begin{enumerate}
\item Select the most probable sky location fulfilling the IACT observation conditions (e.g. zenith angle range, dark time, etc.) following the definitions which are provided in the next subsections. 
\item Schedule observation for this direction at $T_0$ with a duration $\Delta$t.
\item Mask a circular sky region representing the effective IACT field-of-view around that region. Note that this condition allows for the overlap of observing region whenever it is beneficial for the total probability coverage maximization.
\item Using the modified visibility window $T_i =T_0 +  i \cdot \Delta$t, where $i$ is the observation number, and the iteratively masked skymap, steps 1-3 are repeated until $\gamma$-ray emission is detected by the real-time analysis, the covered probability for the next observations is insignificant or the allocated observation time is used.   
\end{enumerate}

In the following we describe the various options available for the crucial first step, {\it aiming at selecting the most probable sky location}, in this procedure. 

Independent of the option chosen, the scheduling should prioritize sensitivity to low $\gamma$-ray energies (cf. Sec.~\ref{sec:GWfollowup}). Therefore, a module is included in the scheduling algorithms allowing to favor low zenith angle observations. This prioritisation is performed via a scan of different maximum zenith angles allowed for each pointing and using a weight that relates the expected gain in energy threshold to the less optimal coverage of the GW uncertainty region. 

\subsection{2D Scheduling Algorithms}\label{sec:2Dalgorithms}
The most straightforward approach to the scheduling problem is to use the two-dimensional localization probability provided with the GW alerts, $\rho_i$, which represents the posterior probability that the source is contained inside pixel $i$. In the following we will refer to this quantity as $P_{GW}$. The scheduling algorithms determine the pointing pattern of the telescopes by trying to cover most of the GW localization region, i.e. trying to maximise $P_{GW}$ within the FoV of the telescope.\\

\underline{\texttt{Best-pixel} algorithm}\\
This strategy is based on pointing observations according to the selection of individual high probability pixels $P^i_{GW} = \rho_{i}$ in the HealPix skymap provided by the GW instruments, which correspond to coordinates (RA$_i$, Dec$_i$). Each pointing is centered on the pixel with the highest probability value. The region around the pixel falling into the region defined by a circle of radius $r$ = r$_{FoV}$ is assumed as covered. The algorithm selects regions based on a single-pixel probability at whose coordinates the observation is centered, so in a large number of cases, the coordinates for the observations for $N_{obs}>$1 will be selected to be next to the edge of the already covered region in previous observations. Due to this effect, the resulting pointing pattern is typically characterized by a significant degree of overlap between the individual pointings. 
This is in general an undesired scenario when searching for a transient source in a large sky region with a rather homogeneous localisation probability distribution throughout the uncertainty region, so an optimised algorithm is described in the following, which matches the capabilities of an intermediate FoV telescope.\\ 

\underline{\texttt{PGW-in-FoV} algorithm}\\
Instead of selecting the pixel with the highest probability, one can investigate the most probable region. The pointing is chosen for the coordinates where $P^{\text{FoV}}_{\text{GW}}$, defined as 


\begin{equation}
P^{\text{FoV}}_{\text{GW}}=\int^{2\pi}_{0}\int_0^{r_{\text{FoV}}}\rho(r,\phi) \,\mathrm{d}r \mathrm{d}\phi,
\label{eq:PGWFOV}
\end{equation}

reaches its maximum.\\

\begin{figure}
  \centering
  \begin{minipage}[b]{0.23\textwidth}
    \includegraphics[width=\textwidth]{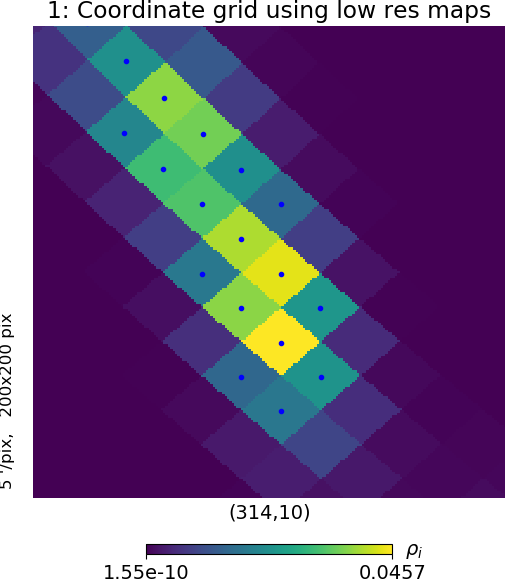}
  \end{minipage}
  \hfill
  \begin{minipage}[b]{0.23\textwidth}
    \includegraphics[width=\textwidth]{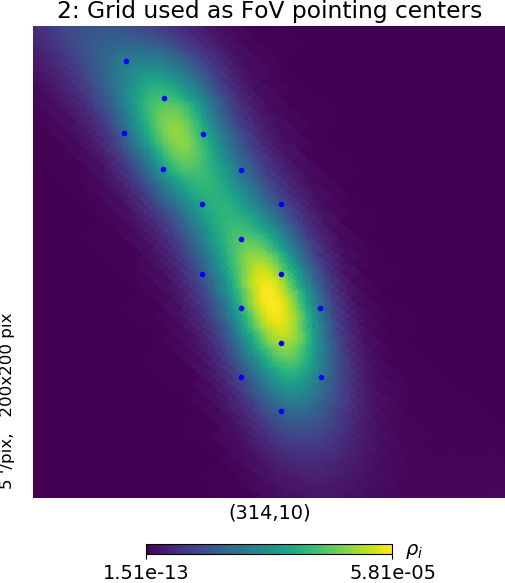}
    \end{minipage}
      \hfill
     \begin{minipage}[b]{0.23\textwidth}
    \includegraphics[width=\textwidth]{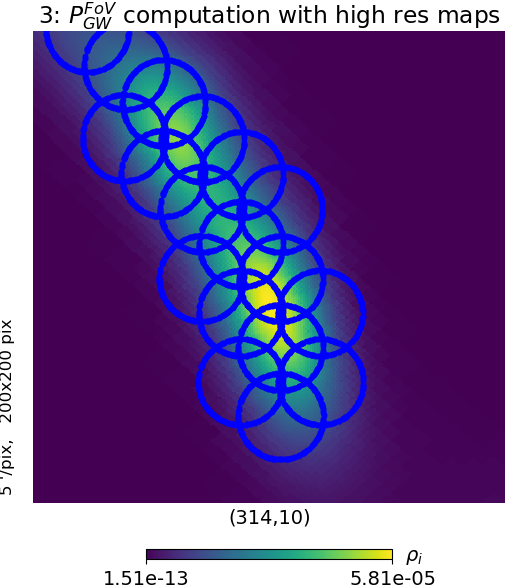}
  \end{minipage}
  \hfill
  \begin{minipage}[b]{0.23\textwidth}
    \includegraphics[width=\textwidth]{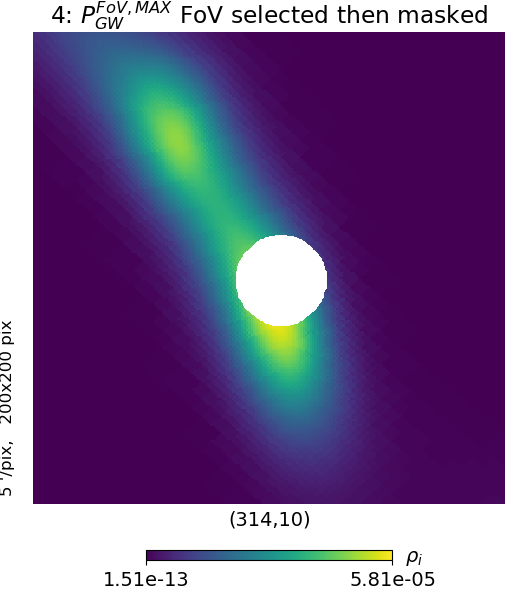}
  \end{minipage}
  \caption{Graphical representation of the \texttt{PGW-in-FoV} algorithm steps to compute the best observable position for one window. These steps are repeated to determine the best position to observe for each observation window. The \textit{initial} map for S190728q is used as an example. For representation purposes the region enclosed in the  50\% localisation uncertainty and $N_{side}$ =32 are chosen for the construction of the low resolution coordinate grid. The blue dots represent the grid of coordinates of the IACT FoVs inside which $P^{\text{FoV}}_{\text{GW}}$ will be calculated. The blue circles are the considered FoVs . The white region is masked.}
  \label{fig:PGW_steps}
\end{figure}

In order to reduce the computation time of Eq.~\ref{eq:PGWFOV}, an additional feature has been added to the algorithm which profits from the fast re-pixelizations of the HEALPix maps. The method is based on the use of an auxiliary probability skymap which is rebinned in a way that the bin areas are close to the FoV$_\text{{IACT}}$. The centers of the new bins define a grid of coordinates as shown in Fig.~\ref{fig:PGW_steps} and represent the center of the H.E.S.S. FoVs inside which $P^{\text{FoV}}_{\text{GW}}$ is calculated. Thus, the algorithm considers in parallel two skymaps: a low resolution one, which is used as a grid of coordinates to scan the GW localisation region then a high resolution one, in order to obtain a good computation of $P^{\text{FoV}}_{\text{GW}}$. Then the highest probability sky region that fullfils $P^{\text{FoV},i}_{\text{GW}}=P^{\text{FoV},MAX}_{\text{GW}}$ is chosen to be observed and masked for the next computation corresponding to the following visibility window. The same steps are repeated for the following window taking into consideration changes in observation conditions as mentioned above and previously masked regions in the skymap. The scheduling algorithm performance is optimised using this feature; the computation time which increases linearly with the number of tested pixels, as well as the accuracy of the calculation which scales with the difference between the resolution of the map and the size of the FoV$_\text{{IACT}}$.

The parallel use of two different resolution skymaps allows to incorporate an extra feature to the algorithm. Using the low-resolution skymap, a certain percentage of the localization uncertainty can be determined by selecting the pixels from the coordinate grid enclosed in the x\% probability uncertainty region in a reasonable amount of time. This modification is included in all the algorithms that use the parallelisation of two different resolution skymaps. In H.E.S.S., we adopt a 90\% value, which following the definition of the probability, implies that potentially we would not cover 10\% of the events. Nevertheless, these regions are in most cases \textit{indirectly} covered due to the intermediate FoV of IACT telescopes.

\subsection{3D Scheduling Algorithms}\label{sec:3Dalgorithms}
The matter distribution in the local universe is inhomogeneous and can be traced by the distribution of galaxies. Assuming that BNS systems form predominantly within galaxies, we can exploit the inhomogeneities to improve the search for the remnants of BNS mergers. Following this reasoning, the search region can be reduced, and the chances of detecting the EM counterpart can be increased by convolving the GW localization region with the galaxies that could plausibly host such cataclysmic events. The prioritisation of galaxies within the GW uncertainty region can potentially decrease the number of observations needed to cover the most plausible source locations as well as limit the number of false positives~\citep{abadie2012implementation}. Although the link between the BNS merger rate and galaxy properties, such as mass, luminosity and star formation rate, are not yet clearly established, the convolution of the 3D-localisation uncertainty by the distribution of matter plays an unambiguous role in decreasing the uncertainty regions~\citep{singer2016going}.\\

Following~\citep{singer2016going}, a posterior probability \textit{volume} can be defined, which represents the probability that the source is located within a pixel \textit{i}, corresponding to the coordinates (RA$_i$, Dec$_i$) at a distance [$z$,$z$+d$z$], from which the probability density per unit volume normalised to unity in Cartesian coordinates is given as: 

\begin{equation}
\frac{\mathrm{d}P}{\mathrm{d}V} = \rho_i \frac{N_\mathrm{pix}}{4\pi} \frac{\hat{N_i}}{\sqrt{2\pi}\hat{\sigma}_i}\text{exp}\bigg[ -\frac{(z-\hat{\mu_i})^2}{2\hat{\sigma}_i^2} \bigg] 
\label{eq3}
\end{equation}

where $\hat{\mu}_i$ = $\hat{\mu}(n_i)$, $\hat{\sigma}_i$ = $\hat{\sigma}(n_i)$, $\hat{N}_i$ = $\hat{N}(n_i)$ refer to the location parameter, the scale, and the normalization. The convolution of the 3D posterior probability distribution of the localization of a GW and a three dimensional distribution of \text{potential} hosts in the Universe, define a new normalized probability which in the following is referred as $P_{\text{GWxGAL}}$, which is defined as: 

\begin{equation}
P^i_{\text{GWxGAL}}=\frac{\mathrm{d}P^i/\mathrm{d}V}{\sum_j \mathrm{d}P^j/\mathrm{d}V}
\label{eq4}
\end{equation}

where $\sum_i P^i_{\text{GWxGAL}}$ = 1. Using Eq.~\ref{eq4}, we developed three optimized strategies for GW events occurring at distances for which \textit{reasonably} complete galaxy catalogs are available. Further details on this important consideration are given in Sec.~\ref{sec:galaxycats}.\\

\underline{\texttt{Best-galaxy} algorithm}\\
The coordinates of the most promising pointing are chosen according to the selection of individual high probability galaxies and the observation of those one-by-one. In each iteration, the galaxy with the highest probability, $P^i_{\text{GWxGAL}}$ guides the observations and the galaxies included in the region defined by the FoV of the telescope are indirectly observed. The motivation of such algorithm is the trade-off between the speed of computation, which is a key point for transient searches, and the astrophysical motivation. However, the observation schedule resulting from this technique can as well present important overlapping of covered regions, as it was the case for the \texttt{Best-pixel} algorithm. Furthermore, although such galaxy-targeted searches can be very performant for small FoV instruments, like optical and X-ray telescopes~\citep{gehrels2016galaxy}, the relatively large FoVs of IACTs motivate a further step in the selection of the observation coordinates.\\

\underline{\texttt{Galaxies-in-FoV} algorithm}\\
Medium-FoV experiments like IACTs do benefit from the integration of probability regions in the sky following Eq.~\ref{EqGal}, where the goal is not only to cover the maximum probability region but also to target galaxy clusters instead of individual galaxies. In this approach, we define $P^{\text{FoV}}_{\text{GWxGAL}}$ as:


\begin{equation}
P^{\text{FoV}}_{\text{GWxGAL}}=\int^{2\pi}_{0}\int_0^{r_{\text{FoV}}}P^i_{\text{GWxGAL}}(r,\phi)\, \mathrm{d}r \mathrm{d}\phi.
\label{EqGal}
\end{equation}

The galaxies are here taken as positional seeds at the center of the FoV$_{\text{IACT}}$ and the total probability of all individual galaxies contained in the FoV is computed for a large enough number of galaxies. The highest probability sky field, which fulfills $P^{\text{FoV},i}_{\text{GWxGAL}}=P^{\text{FoV},MAX}_{\text{GWxGAL}}$ is chosen \footnote{\texttt{Galaxies-in-FoV} is referred to as \texttt{PGalinFoV},  $P^{\text{FoV,MAX}}_{\text{GWxGAL}}$ as $P_{\text{GAL}}$ and $P^{\text{FoV,MAX}}_{\text{GW}}$ as $P_{\text{GW}}$ in the following.} to be observed during the given window, and is masked for the following iterations.\\

\begin{figure}
  \centering
  \begin{minipage}[b]{0.23\textwidth}
    \includegraphics[width=\textwidth]{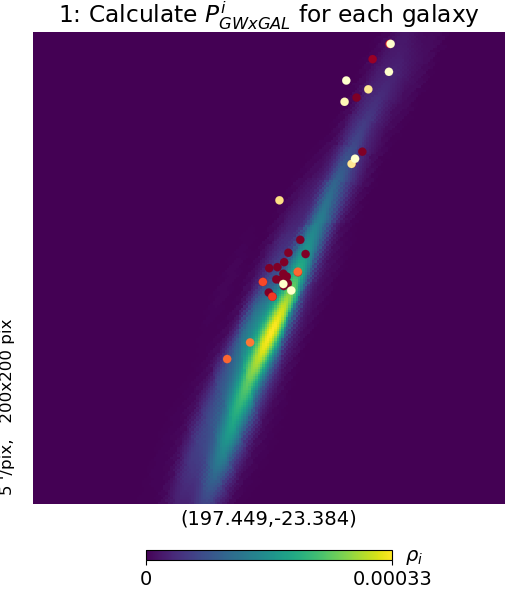}
  \end{minipage}
  \hfill
  \begin{minipage}[b]{0.23\textwidth}
    \includegraphics[width=\textwidth]{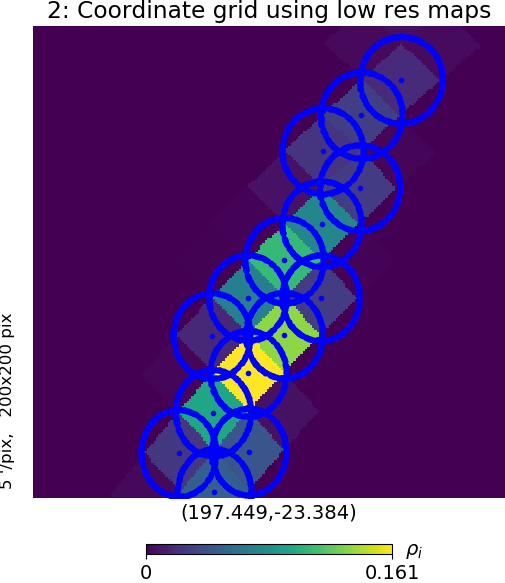}
    \end{minipage}
      \hfill
     \begin{minipage}[b]{0.23\textwidth}
    \includegraphics[width=\textwidth]{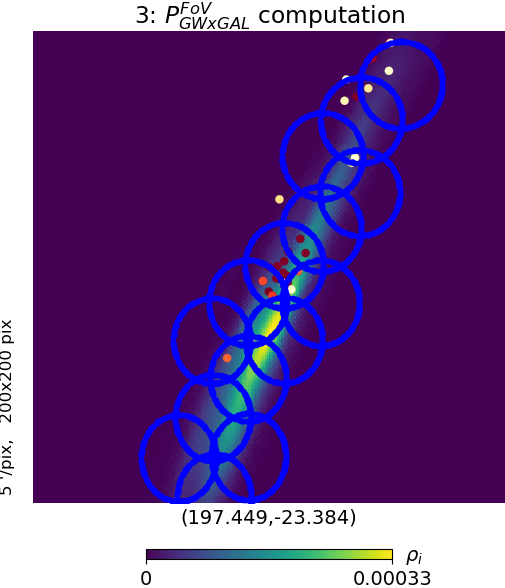}
  \end{minipage}
  \hfill
  \begin{minipage}[b]{0.23\textwidth}
    \includegraphics[width=\textwidth]{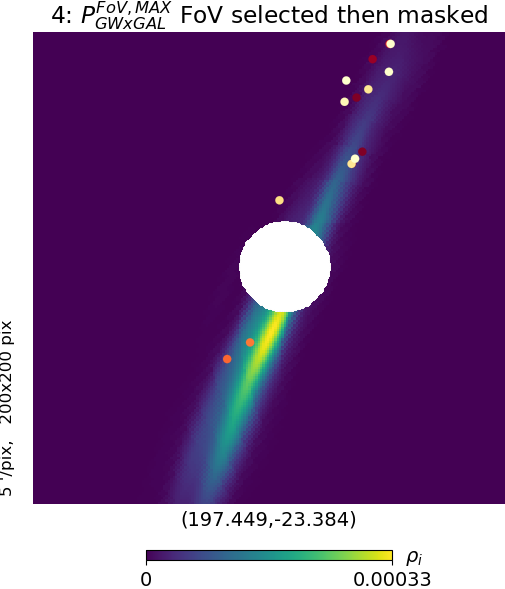}
  \end{minipage}
  \caption{Graphical representation of the \texttt{PGalinFoV-PixRegion} algorithm steps to compute the best observable position for one window. These steps are repeated to determine the best position to observe for each observation window. The \textit{updated} map for GW170817 is used. For representation purposes the region enclosed in the  90\% localisation uncertainty and $N_{side}$ = 32 are chosen for the construction of the low resolution coordinate grid. The colored dots represent the galaxies with the highest $P^i_{GWxGAL}$  in the region on the $\textit{YlOrRd}$ color scale. The blue dots represent the grid of coordinates of the IACT FoVs inside which $P^{\text{FoV},i}_{\text{GWxGAL}}$ will be calculated. The blue circles are the considered FoVs . The white region is masked.}
  \label{fig:PGal_PR_steps}
\end{figure}

\underline{\texttt{PGalinFoV-PixRegion} algorithm}\\
The number of galaxies enclosed in the probability volume increases drastically with increasing GW localisation uncertainty regions, requiring large number of operations to be performed to determine $P^{\text{FoV},MAX}_{\text{GWxGAL}}$. The \texttt{PGalinFoV-PixRegion} algorithm has been developed to address this challenge. The \texttt{Galaxies-in-FoV} algorithm is updated with one of the main features of the \texttt{PGW-in-FoV} algorithm: the parallel use of two different resolution skymaps. The \texttt{PGalinFoV-PixRegion} algorithm uses a low resolution rebinned skymap (typically with $N_{side}$ = 64) as a coordinate grid for the pointing seeds like the \texttt{PGW-in-FoV} algorithm and chooses $P^{\text{FoV},MAX}_{\text{GWxGAL}}$ which is computed from a much higher resolution map convoluted with the galaxy catalog as explained in Fig.~\ref{fig:PGal_PR_steps}. For 3D searches we use an enclosed region of 99\% (instead of 90\%) to make sure that we cover all galaxies at the edges of the GW skymaps.\\


\subsubsection{Galaxy catalogs}\label{sec:galaxycats}
In order to use the cross-correlation introduced in the previous section, two catalogs of galaxies have been considered: The Census of the Local Universe (CLU) catalog~\citep{cook2017census} and 
The Galaxy List for the Advanced Detector Era (GLADE) catalog ~\citep{dalya2018glade}. The latter is currently being used in the H.E.S.S. GW program due to its easy availability and continuous updating.\\

The GLADE catalog has been built by cross-matching five, non-independent astronomical catalogs, including galaxies and quasars. The catalog completeness has been assessed by the authors of the catalog in terms of cumulative blue luminosity outside the Galactic plane. It is found to be fully complete up to $d_L=37^{+3}_{-4}$ Mpc, and it has a completeness of $\sim$ 61, $\sim$ 54, $\sim$ 48 percent within the maximal value of single-detector BNS ranges for aLIGO during O2, O3, and design sensitivity, respectively ~\citep{dalya2018glade}.\\

Besides the limitation due to completeness, Another limitation that has to be considered for the use of galaxy catalogs is the region which corresponds to the Galactic plane (GP). Experimental difficulties to perform galaxy targeting observations with the GP in the line-of-sight lead to a significantly lower number density of objects in this region. The use of a galaxy based approach for GW follow-ups in the vicinity of the GP would therefore be inefficient and biased. Within the H.E.S.S. GW follow-up program we define an \textit{avoidance zone} around the GP, where the region has the geometrical form of a rhombus, centered in the Galactic Center. It represents $\sim$4\% of the sky. For GW events whose maximum value of the localization confidence regions falls inside the avoidance zone, a 2D scheduling approach is selected.\\

A further modification on the catalog is made regarding objects whose luminosity distance is far greater than the fiducial BNS horizon for the detections made by LVC collaboration. This pre-processing step removes remote extragalactic objects (mainly AGNs) and reduces the catalog size to a quarter of the initial number thus increasing  the processing speed.

\subsection{Performance estimates and comparisons}
\begin{figure}
  \begin{minipage}[t]{0.49\textwidth}
    \includegraphics[width=\textwidth]{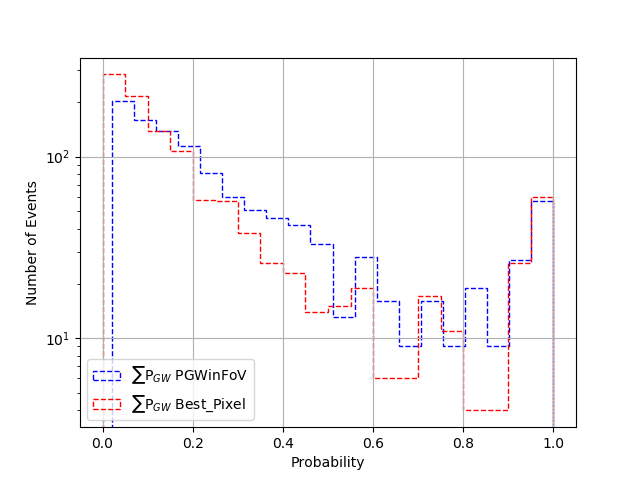}
    \caption{Total P$_{GW}$ simulated coverage distribution for the \texttt{PGW-in-FoV} and \texttt{Best-pixel} algorithms.}
   \label{fig:ProbDist_PGW_2D}
  \end{minipage}
  \hfill
  \begin{minipage}[t]{0.49\textwidth}
    \includegraphics[width=\textwidth]{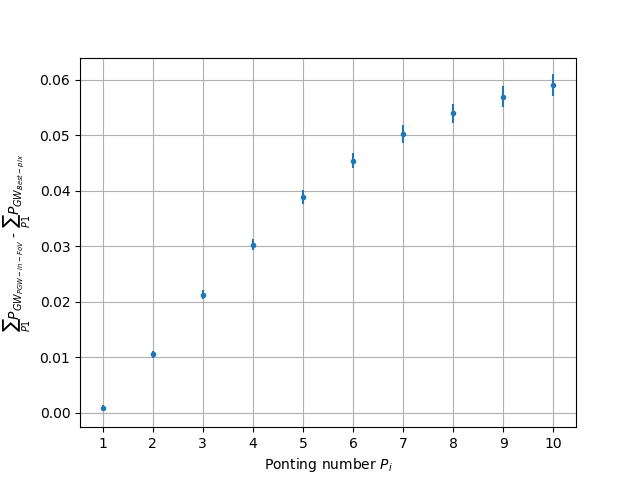}
    \caption{Difference of the cumulative P$_{GW}$ per pointing between \texttt{PGW-in-FoV} and \texttt{Best-pixel}}.
   \label{fig:ProbPerPointing}
  \end{minipage}
\end{figure}

In this section we present the tests results we use to determine which of the algorithms described in this section is better suited for H.E.S.S. For 2D coverage we consider \texttt{Best-pixel} and \texttt{PGW-in-FoV}. \texttt{PGW-in-FoV} relies on a time and space dynamic tiling that tracks and adapts to the motion of the GW localisation region in the sky during the night. Like the \textit{greedy} strategy used by the Zwicky Transient Facility~\citep{ZTF_tiling}, it considers the integrated probability inside the FoV. \texttt{Best-pixel} is a more straight forward approach that does not require actual tiling. In order to asses and compare their performance, we test them on 250 simulated BAYESTAR GW maps from~\citep{singer2016going}. We inject the event at random times throughout a whole year. For each map an optimized observation schedule following both algorithms is derived for 10 different times which gives a total of 2500 trials for each algorithm. Considering the necessary observation conditions for H.E.S.S. and imposing a minimum requirement coverage of P$_{GW}$ = 2\% per observation we compare in Fig.~\ref{fig:ProbDist_PGW_2D} the total P$_{GW}$ coverage that can be achieved in the first night of observations. Only simulations where at least one observation is scheduled are taken in consideration. A slightly larger P$_{GW}$ is covered by \texttt{PGW-in-FoV}. The difference in the cumulative P$_{GW}$ (for one night) for the two algorithms is computed up to 10 pointings for each simulation. The mean (and error on the mean) of these values for the 2500 simulations are shown in Fig.~\ref{fig:ProbPerPointing}. The coverage per pointing is less efficient for \texttt{Best-pixel} as the pointing number increases during a follow-up which is due to the FoVs overlap. This causes the \texttt{Best-pixel} strategy to struggle with achieving the minimum coverage requirement per pointing in the case of extended GW maps (large localisation regions) since fewer effective pointings will pass the minimum coverage requirement cut resulting in a smaller total P$_{GW}$ coverage. \texttt{PGW-in-FoV} starts to improve coverage immediately after the first observation and this improvement reaches up to to 6\% within 10 observation runs on average, knowing that in some cases it is significantly higher. It is important to mention here that not all simulations will reach 10 scheduled pointing per night. Overlap could be reduced by significantly lowering the GW resolutions for the \texttt{Best-pixel} algorithm but this comes at the cost of accuracy in probability computation. Therefore we conclude that \texttt{PGW-in-FoV} performance is superior to \texttt{Best-pixel}.\\

\begin{figure}[!htb]
  \centering
  \begin{minipage}[b]{0.49\textwidth}
    \includegraphics[width=\textwidth]{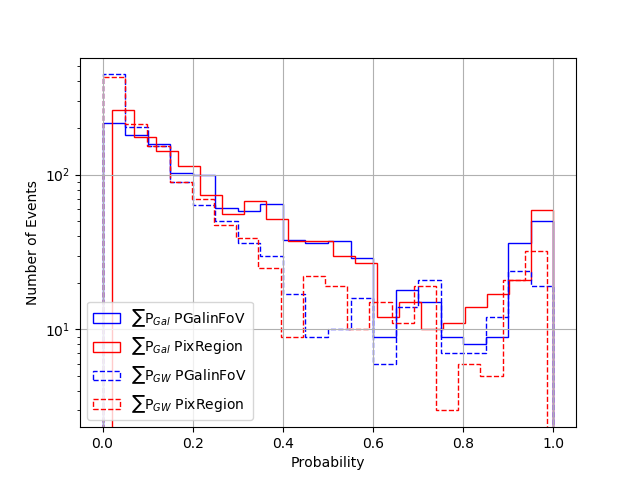}
    \caption{Total P$_{GAL}$ and P$_{GW}$ simulated coverage distributions for \texttt{PGalinFoV} and \texttt{PGalinFoV-PixRegion}.}
   \label{fig1:ProbDist_PGal_PixRegion}
  \end{minipage}
    \hfill
\begin{minipage}[b]{0.49\textwidth}
    \includegraphics[width=\textwidth]{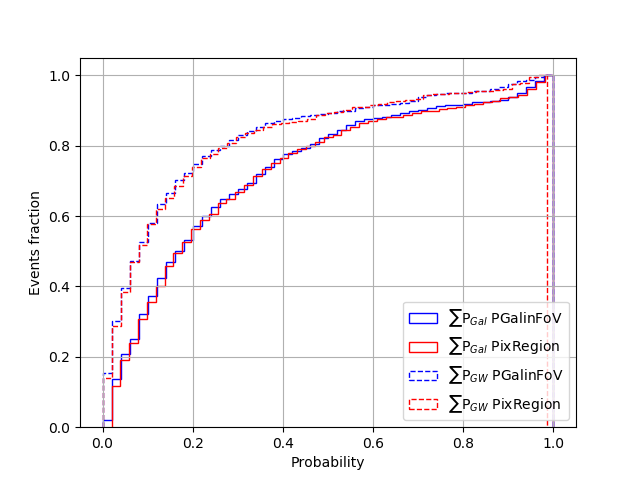}
    \caption{Total P$_{GAL}$ and P$_{GW}$ simulated coverage cumulative distributions for \texttt{PGalinFoV} and \texttt{PGalinFoV-PixRegion}.}
   \label{fig2:ProbCum_PGal_PixRegion}
  \end{minipage}
\end{figure}

The 3D algorithms were developed to provide the best possible coverage efficiency of the most promising nearby GW events. The galaxy-targeted \texttt{Best-galaxy} is typically used by small FoV instruments like the GRANDMA collaboration telescopes with $FoV < 1 deg^{2}$~\citep{GRANDMA_GW} and Magellan~\citep{Magellan_GW}. The \texttt{PGalinFoV} and \texttt{PGalinFoV-PixRegion} algorithms are FoV-targeted and are best adapted for medium and large FoV instruments in order to take advantage of the relatively large FoV. FoV-targeted search methods are used by the GRANDMA telescopes with $FoV > 1 deg^{2}$ for GW counterpart searches~\citep{GRANDMA_GW} and ASCAP~\citep{ASKAP_GW}. Hence we consider the \texttt{PGalinFoV} and \texttt{PGalinFoV-PixRegion} algorithms for the H.E.S.S. 3D searches. We test their performance with the same procedure mentioned above for the 2D algorithms. Although the 3D scheduling is based on P$_{GAL}$ values, we also calculate the P$_{GW}$ coverage for each simulated observation. 
To asses the efficiency of \texttt{PGalinFov} and \texttt{PGalinFov-PixRegion} we show  the distibution of the total P$_{GAL}$ and P$_{GW}$ coverage that can be achieved in the first night of observations while taking into account the necessary observation conditions for H.E.S.S. in Fig.~\ref{fig1:ProbDist_PGal_PixRegion} and their cumulative distribution in Fig.~\ref{fig2:ProbCum_PGal_PixRegion}. Both figures show that the coverage efficiency for P$_{GAL}$ and P$_{GW}$ of both algorithms is comparable.\\

In practice the scheduling framework is typically started by loading the telescope parameters, the galaxy catalogs and the GW skymap. While the first two steps can be performed at the beginning of the observation period, the download and analysis of the GW map happens on an event-by-event basis and can only be performed after the alert has been issued by LVC. Any potential map rebinning takes place of course after it is loaded so the loading time poses a limitation to our time saving measures. The skymap is then correlated with the catalog of galaxies following Eq.~\ref{eq4}. We then define the available observations windows according to current and future observational conditions before calculating the optimal pointing for each of them. The \texttt{PGalinFoV-PixRegion} algorithm requires an additional step for the determination of the seed coordinate grid prior to the final probability calculation.

\begin{figure}[!htb]
  \centering
  \begin{minipage}[b]{0.49\textwidth}
    \includegraphics[width=\textwidth]{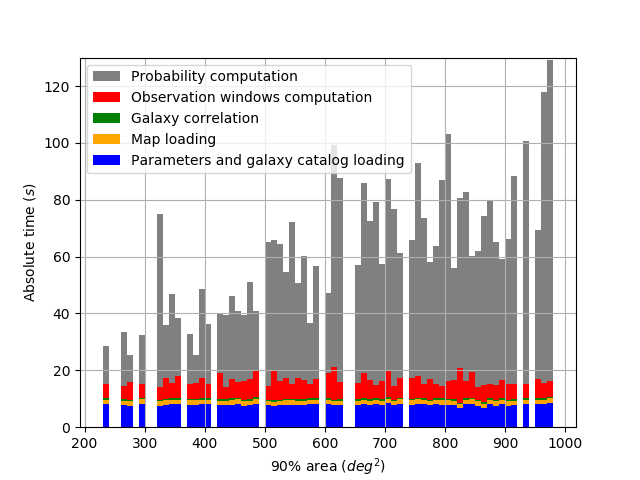}
    \caption{Absolute time for the computation of each step of the PGalinFov algorithm for $N_{side}$ = 512 maps up to an area of $1000~\mathrm{deg}^2$.}
  \label{fig:PGalinFoV_Times}
  \end{minipage}
    \hfill
  \begin{minipage}[b]{0.49\textwidth}
    \includegraphics[width=\textwidth]{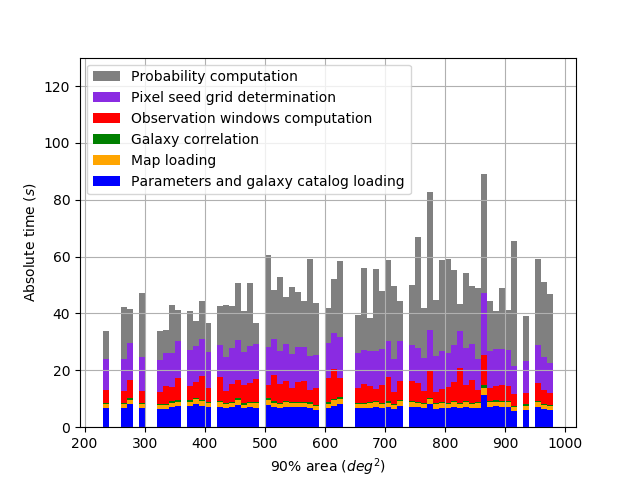}
    \caption{Absolute time for the computation of each step of the PGalinFov-PixRegion algorithm for $N_{side}$ = 512 maps up to an area of $1000~\mathrm{deg}^2$.}
    \label{fig:PixRegioninFoV_Times}
    \end{minipage}
\end{figure}

Fig.~\ref{fig:PGalinFoV_Times} shows the time required by each of these steps for the \texttt{PGalinFoV} algorithm as a function of the size of the GW localisation area of the simulated events (defined as the area containing 90\% of the total probability). The time required to load and analyze the GW map depends its $N_{side}$ resolution. It takes on average $\sim$30 seconds to load a high resolution map with $N_{side}$ = 2048, while it takes only $\sim$8 seconds to load a map with resolution $N_{side}$ = 1024. Maps with $N_{side}$ = 512 require in average less than 2 seconds and maps with $N_{side}$ = 256 require less than 1 second to be loaded. Due to limited number of available simulations we here concentrate on the most common maps, those with $N_{side}$ = 512. We note that, unfortunately, high resolution maps with $N_{side}$ = 2048 correspond generally to relatively well localized GW events. As the localisation uncertainty is related to the signal strength as well as the number of participating GW interferometers, these events are rare (only 5 of the 80 events detected during O3 have been distributed with $N_{side}$ = 2048), but are at the same time the promising ones for rapid detections of counterparts which poses as a limitation to our time saving measures.

Fig.~\ref{fig:PGalinFoV_Times} shows that, while the time required for each step in the computation of the schedule is constant as a function of the size of the GW map the time to define the best pointing direction, i.e. the {\textit probability computation} necessary to define each $P_{\text{GAL}}$, increases with the GW localisation uncertainty for \texttt{PGalinFoV}. This is due to the fact that the number of galaxies used as seeds for the computation increases drastically with increasing uncertainty \textit{volume}, thus requiring more calculations to be preformed.

To determine the time required by each step of \texttt{PGalinFoV-PixRegion}, we repeat the previous study for this algorithm. As illustrated in Fig.~\ref{fig:PixRegioninFoV_Times}, the time required for the \textit{probability calculation} defining $P_{\text{GAL}}$ remains stable when the localisation uncertainty increases. This is due to the fact that the number of pixels used in the 2D observational seeds grid is based on a low resolution map and therefore does not increase as drastically as the number of target galaxies within the uncertainty volume. However, the additional step necessary to define the 2D seeds grid adds a considerable amount of time to the overall budget.

We note that the absolute values of the computation time shown here are dependent on the parameters and performance of the machine(s) performing the operation and only the general behavior of the data should be taken into consideration.

In conclusion, the \texttt{PGalinFoV-PixRegion} algorithm allows to cover large maps faster than \texttt{PGalinFoV} which makes it more suitable for the coverage of poorly localised events. On the other hand, \texttt{PGalinFoV} is slightly faster for well localized maps and is therefore the preferred option for the most interesting, high signal-to-noise GW events detected by the full LVC network. While both algorithms are available within the H.E.S.S. GW follow-up framework, we currently use \texttt{PGalinFoV} for  3D scheduling computations.
\section{The H.E.S.S. GW follow-up program}
\label{sec:program}
The GW follow-up algorithms are provided in the GW follow-up plugin integrated into the H.E.S.S. Transients follow-up system introduced in Sec.~\ref{sec:GWfollowup}. This system reacts to all publicly available GW alerts. For each signal detected by LVC, the full sequence of notices is processed to find follow-up opportunities. During the run O3 these included \textit{preliminary}, \textit{initial}, \textit{update} and \textit{retraction} notices. In preparation of the run O4, an Early Warning notice has been added in June 2020~\citep{earlyWarning}.

\begin{figure}
  \centering
\includegraphics[width=0.95\textwidth]{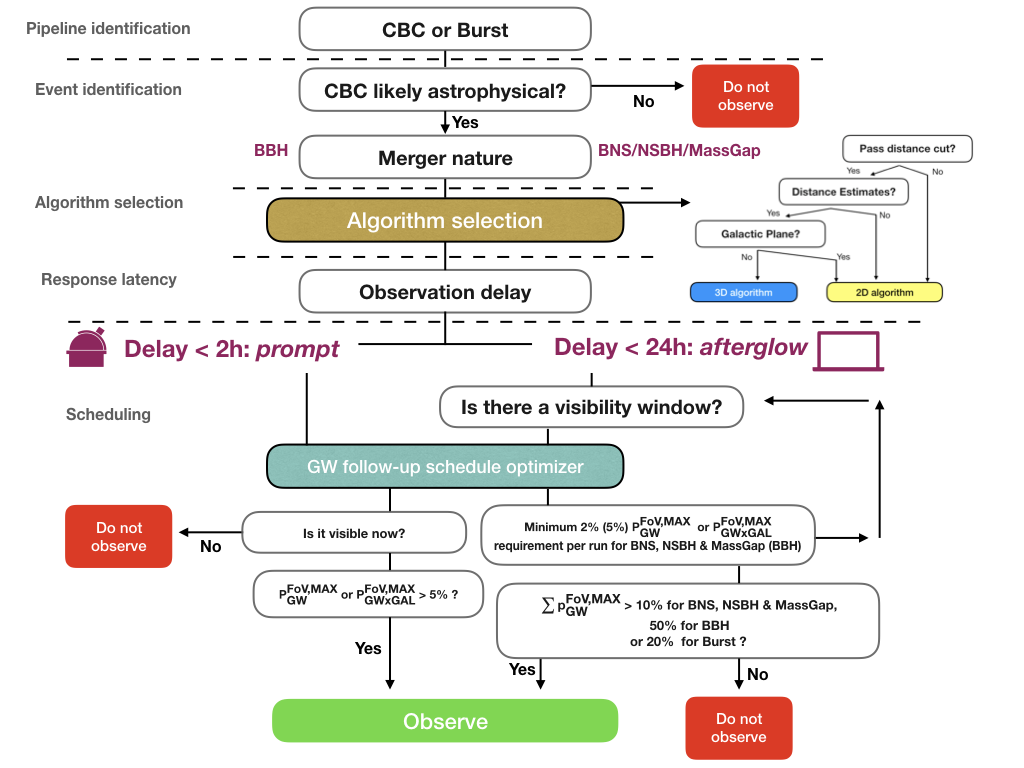}
\caption{Schematic overview of the decision tree used in the automatic response of H.E.S.S to GW events.}
\label{fig:GWVoAlerter}
\end{figure}

The progress of an incoming event throughout the decision tree outlined in Fig.~\ref{fig:GWVoAlerter} is monitored via email alerts at all major decision points. This allows the expert on call to follow all steps remotely. The precise filtering criteria outlined here have been iterated several times during O2 and O3, are subject to further changes and should therefore only be considered indicative.

Upon the arrival of an alert, the H.E.S.S. ToO alert system assigns a science case to it. For GW alerts, three science cases are currently available: BNS which also concerns all CBC mergers involving at least one NS (BNS, NSBH and MassGap), BBH and Burst alerts. Burst events are un-modelled GW events that can be caused by rare but interesting sources like nearby supernovae. Their reconstruction does not include a distance estimate of the event. 
Thew selection is done based on the information provided in the VoEvent message emitted by LVC starting with the GW detection pipeline to determine if it is a Burst or a CBC alert. CBCs are then evaluated based on their probability to be likely astrophysical in origin. At this step, events with high noise probability ($>$ 50\%) that are less likely to be astrophysical are filtered out. 

Depending on the BBH probability parameter in the VoEvent alert message, the alerts are processed either as being emanating from a BBH merger (probability BBH $>$ 50\%), or a merger including at least one neutron star (BNS science case). The system then downloads the corresponding localisation map and proceeds with the selection of the optimal scheduling algorithm, i.e. the choice of a 2D or a 3D strategy. Considering the completeness of the GLADE galaxy catalog (cf~\ref{sec:galaxycats}), only GW events having a mean distance $<$ 150 Mpc and having their GW map hotspot outside the \textit{avoidance zone} are analyzed using a 3D approach. Events lacking distance information or not fulfilling the mentioned criteria are treated with a 2D approach. Only in the case of a 3D coverage the galaxy catalog is loaded. The localisation map is then forwarded to the GW follow-up schedule optimizer that derives suitable obsrvational strategies using the algorithms described in Sec.~\ref{sec:algorithms}.

For events selected for a 2D treatment, the \texttt{PGW-in-FoV} algorithm is used in the GW follow-up schedule optimizer. For events that allow a full 3D analysis, the \texttt{PGalinFoV} algorithm is used due to its faster reaction time to small maps (see Fig.~\ref{fig:PGalinFoV_Times} and~\ref{fig:PixRegioninFoV_Times}). \texttt{PGalinFoV} is also better suited than \texttt{PGalinFoV-PixRegion} for single \textit{prompt} reaction since it does not include the additional time consuming step of computing the 2D pixel observational seeds grid.

For alerts that arrive during observation time and that would therefore allow for prompt observations, a rapid response is of utmost importance. We therefore divided the GW follow-up schedule optimizer into two modules: \textit{prompt} and \textit{afterglow}\footnote{Although the \textit{prompt} and \textit{afterglow} take the names of physical emission phases they are not directly related to them.}. The \textit{prompt} module is available for low latency alerts (\textit{preliminary} and \textit{initial}), the \textit{afterglow} module handles also \textit{updates}. Instead of calculating the entire schedule for the available dark-time period, the \textit{prompt} module only computes the first P$_{GAL}$ or P$_{GW}$ at the time of the arrival of the alert, taking into consideration the visible parts of the sky at that time. The obtained observation direction is then assessed by considering zenith angle, darkness, coverage and time delay conditions. If at the end of the decision making tree all observational criteria are met, observations are forwarded automatically to the data acquisition and slow control system of H.E.S.S. for more information). Without the need for time consuming human interventions the telescopes would automatically stop the current data taking, slew to the derived pointing and start observations. Meanwhile, the \textit{afterglow} module independently computes an entire schedule for the full available observation time of the current night by selecting the most probable position for each possible observation window in the iterative way outlined in Sec.~\ref{sec:algorithms}. 

For alerts that arrive outside H.E.S.S. observation time, the \textit{prompt} module will discard immediate automatic observation possibilities and the \textit{afterglow} module will schedule observations for the upcoming night as described above. The observation schedule will be automatically distributed by email to the observers and the people involved allowing human vetting if needed. At night the observations are performed manually or added to the automatic observation scheduler.

In addition to the scientific decisions described above, an alert has to pass H.E.S.S.-related operational cuts in order to be followed. In general, the aim of H.E.S.S. follow-up observations is to detect or constrain VHE emission associated to the observed GW events. Given that mergers including at least one neutron star are the most promising to show electromagnetic emission, we implemented rather loose filtering cuts to allow for exploratory searches aiming at the detection of a VHE counterpart associated to the GW event. The criteria within the H.E.S.S. GW program require a coverage of at least 10\% of the localisation uncertainty region 
to trigger follow-up observations of a BNS, NSBH or MassGap event. This minimum coverage has to be reached within 24h of the detected GW event. These cuts automatically allow the follow-up of the most interesting events with good localisation and filter out the ones with large localisation uncertainties. 

The aim of the follow-up in BBH mergers, for which electromagnetic emission is less probable, is mainly to constrain the EM emission. This assumption leads to more stringent requirements on coverage achievable with the follow-up observations in order to be able to provide upper limits for most of the possible localisation/emission region. The minimum requirement used within H.E.S.S. for BBH follow-up is 50\% of the GW localisation map. We note that during the past observation runs, all BBH events were detected at large distances exceeding the completeness limits applied for the use of a galaxy catalog. Consequently all BBH events detected so far were analysed with a 2D algorithm.

Burst alerts fall between the two CBC categories and a minimum coverage threshold of 20\% is chosen for their potential promising interest. Alerts that do not pass the filtering cuts are discarded.

In general  a minimum requirement (2 - 5\%) coverage per observation (P$_{GAL}$ or P$_{GW}$) is applied for all GW alerts. For events arriving during the night and having observational delays $<$ 2h, the minimum GW coverage required in order to trigger an automatic \textit{prompt} observation on a position in the sky without waiting for the full schedule to be computed has to be $>$ 5\% for the first pointing.

\begin{figure}
  \centering
  \begin{minipage}[height=0.21\textheight]{0.31\textwidth}
    \includegraphics[height=0.21\textheight]{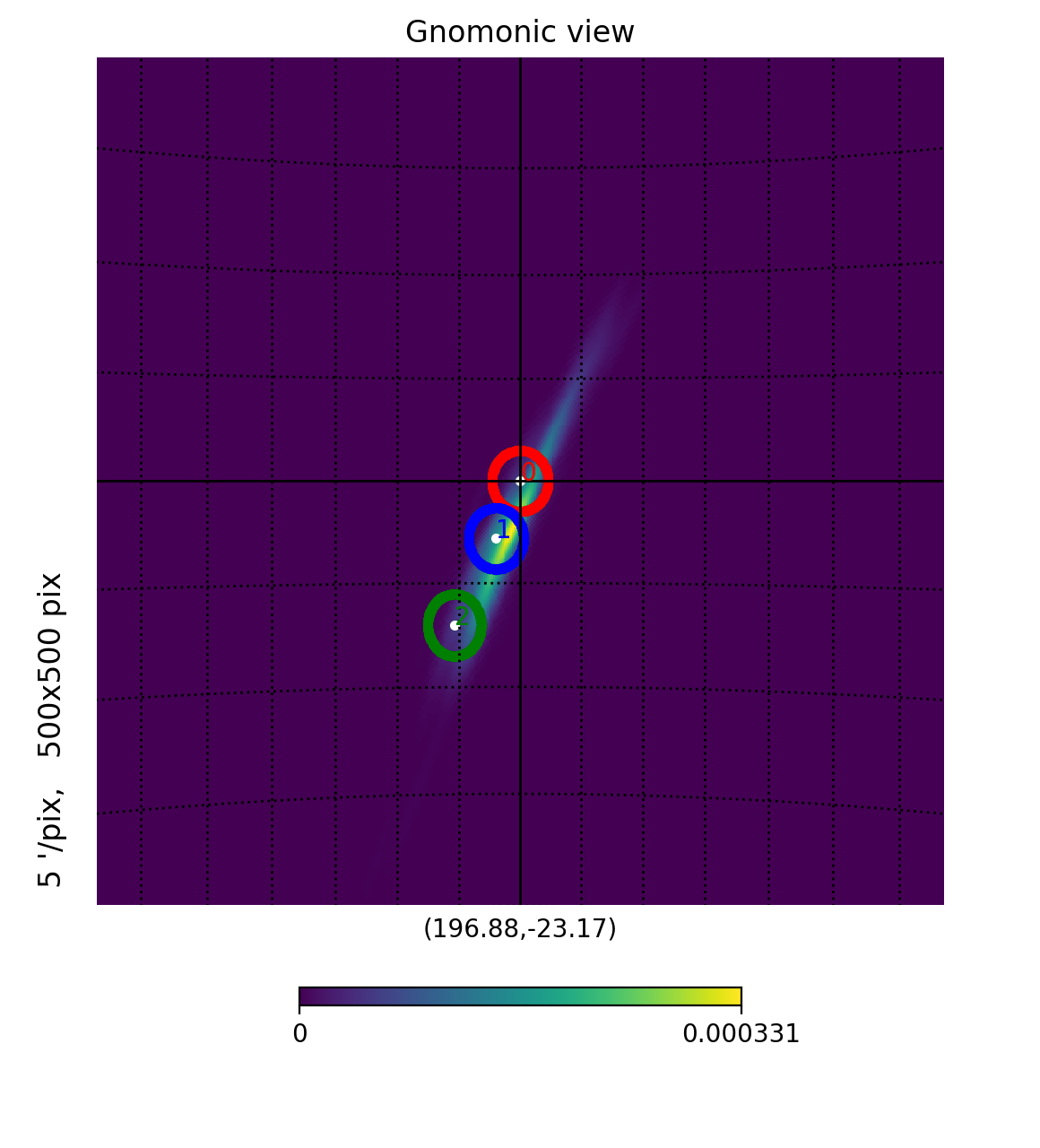}
  \label{fig:GW170814_HESS}
  \end{minipage}
  \begin{minipage}[height=0.21\textheight]{0.62\textwidth}
    \includegraphics[height=0.21\textheight]{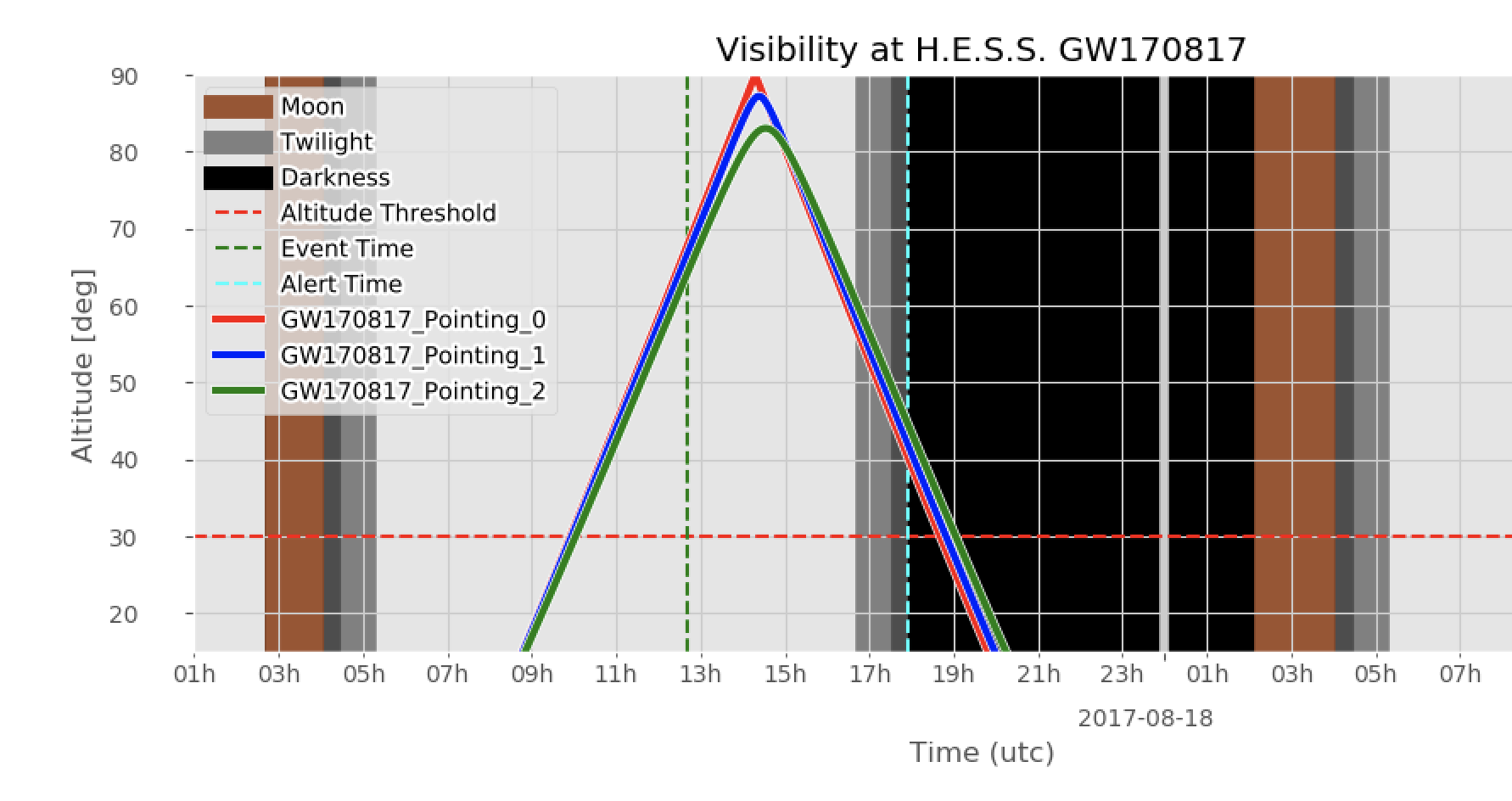}
    \end{minipage}
      \caption{Example of the visual aid provided by the GW follow-up framework for the case of GW170817. On the \textbf{right} the scheduled pointings are superimposed on the GW map with their respective number and on the \textbf{left} the zenith angle evolution of their corresponding position throughout the night is shown.}
  \label{fig:GW170817_graphics}
\end{figure}

The output produced by the implemented GW follow-up tools contains for each observational position the best time to observe and the available observational window throughout the night. This allows for some flexibility in the observations in case the best time could not be respected. All pointings are ranked according to their priority taking into account the achieved coverage and the zenith angle. The observers are presented with both a table like Tab.~\ref{tab:GW170817_schedule} summarizing the proposed scheduling as well as with graphics that illustrate for example the zenith angle evolution of all scheduled positions as illustrated in Fig.~\ref{fig:GW170817_graphics}.

\begin{table}
    \centering
    \begin{tabular}{cccccc}
    \toprule
    Start time &  Ra & Dec & PGAL & Observation window & Priority\\ 
    \midrule
    2017-08-17 17:59 & 196.88 & -23.17 & 0.72 & 2017-08-17 17:55 $\rightarrow$ 2017-08-17 18:39 & 0 \\
    2017-08-17 18:27 & 198.19 & -25.98 &0.16 & 2017-08-17 17:55 $\rightarrow$ 2017-08-17 18:48 & 1 \\
    2017-08-17 18:56 & 200.57 & -30.15 & 0.05 & 2017-08-17 17:55 $\rightarrow$ 2017-08-17 19:01 & 2 \\
    \bottomrule
    \end{tabular}
    \caption{Example of the observation schedule of the GW170817 follow-up. The priority of the pointings is higher in ascending order.}
    \label{tab:GW170817_schedule}
\end{table}

Experts on call are assigned to assist the observers. Their responsibility is to monitor and modulate the automatic response if needed. Offline tools are developed for rapid human intervention. In the case of a GW map {\it update} that arrives after GW ToO observations have already started, a new schedule is automatically computed. These {\it update} maps can also be correlated with external triggers (see RAVEN in~\citep{OpenLVEM}). The expert on call has the ability to  modify this schedule to take into consideration previous observations occuring during the night. Additional tools are being developed in order to monitor the detection of transients and summarize information of known $\gamma$-ray sources in the GW localisation area.

In case of a signal with a significance of more than 5 $\sigma$ is found by the real-time analysis, the observers will re-observe the corresponding position after making sure that the signal does not originate from known VHE sources.

\begin{figure}
  \centering
\includegraphics[width=10cm]{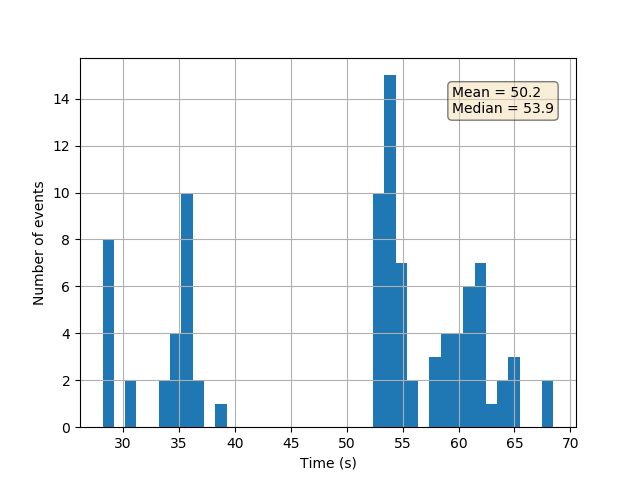}
\caption{The H.E.S.S. \textit{prompt} 3D response time to GW Mock alerts from July to October 2019 excluding telescope slewing time. Only alerts passing all filtering requirements are considered.}
\label{fig:hess_prompt}
\end{figure}

Due to the lack of immediate, \textit{prompt} H.E.S.S. GW follow-up during O2 and O3, we used simulated,\ {\it Mock} alerts in order to quantify the speed of the response of the H.E.S.S. alert system to GW alerts. We select {\it preliminary} and {\it initial} alerts on simulated nearby BNS events that arrive during the night and only consider alerts that pass all observation filtering criteria with a 3D coverage. The results presented in Fig.~\ref{fig:hess_prompt} show that the H.E.S.S. average response time is $\sim$ 50 seconds. This represents the time needed for all steps to be executed in the \textit{prompt} decision tree shown in Fig.~\ref{fig:GWVoAlerter} since the reception of the alert to the distribution of the observational schedule to the shifters, the GW expert team and, most importantly, the DAQ system steering the telescopes. As expected, responses taking more than 50 seconds correspond to events with high resolution localisation maps (e.g. $N_{side}$ = 2048). As explained in Sec.~\ref{sec:algorithms}, these maps have a larger loading and analysis time. Low resolution skymaps with $N_{side}$ $<$ 2048 are clustered at short calculation times on the left side of the histogram (below 40 seconds). The additional telescope slewing time is depending on the distance between the current observations and the target as well as the operation mode. Allowing tracking through zenith in {\it reverse mode}, the 28m H.E.S.S.-II telescope can be on target in less than one minute~\citep{HESS_SLEW}. We can thus expect that H.E.S.S. is able to start observations of a promising GW target often within 1 minute and for the large majority of cases within less than two minutes after receiving the alert.

\section{H.E.S.S. follow-up of GW events}
\label{sec:osbervations}
During O1 and O2 a total of 10 BBH and 1 BNS candidates were reported~\citep{abbott2019gwtc}. In O3, 80 alerts were issued including 24 events that were later classified as noise and retracted, 3 Terrestrial non-retracted events, 52 CBC and 1 unmodeled Burst candidate~\citep{publicAlerts}. Due to the relatively long time delays in the emission of the GW alerts and the large localisation uncertainties during O1 (only the LIGO detectors operated during O1) no observations were conducted by H.E.S.S. in this initial period. The majority of GW events detected by LVC during O2 and O3 could not be followed due to their large localisation regions. Only a few well localised events could not be followed due to weather conditions like in the case of the BBH merger S191204r~\citep{S191204r_initial} and the Burst alert S200114f~\citep{200114f_initial}. Visibility constrains imposed by the full moon did not allow to observe the well localised NSBH merger S190814bv~\citep{S190814bv_initial, GW190814_paper}. However, H.E.S.S. successfully performed follow-up observations of six GW events presented in this section.

The main parameters driving the calculation of the follow-up schedule used during this period are a FoV of radius varying between 1.5 and 2.5 degrees, roughly corresponding to the FoV of the large 28m telescope and the small 12m telescopes respectively, and a maximum zenith angle of 60 degrees. Observations are conducted during a minimum duration of 10 min and with a standard duration for each pointing of 28 minutes\footnote{H.E.S.S. observations are performed with a standard 28 minutes duration due to scientific and technical reasons within the collaboration.}.\\

\subsection{GW follow-up during O2}
The H.E.S.S. GW follow-up scheme has first been applied to real GW data in a technical trial run using the burst alert G284239 identified by LIGO (Hanford + Livingston) although with relatively low significance (e.g. a false alarm rate $<$ 4 per year) at 2017-05-02 22:26:08 UTC. The four observation runs obtained during this ToO were influenced by bad weather conditions but allowed to streamline the follow-up procedures and the communication between offsite experts and the onsite crew that led to the introduction of the full available observation window for each observation to allow flexibility in scheduling when weather and technical difficulties delay observations. The obtained pointing pattern has been made available in GCN~\#21084~\citep{G284239}.\\

\begin{figure}[!htb]
  \centering
  \begin{minipage}[b]{0.44\textwidth}
    \includegraphics[width=\textwidth]{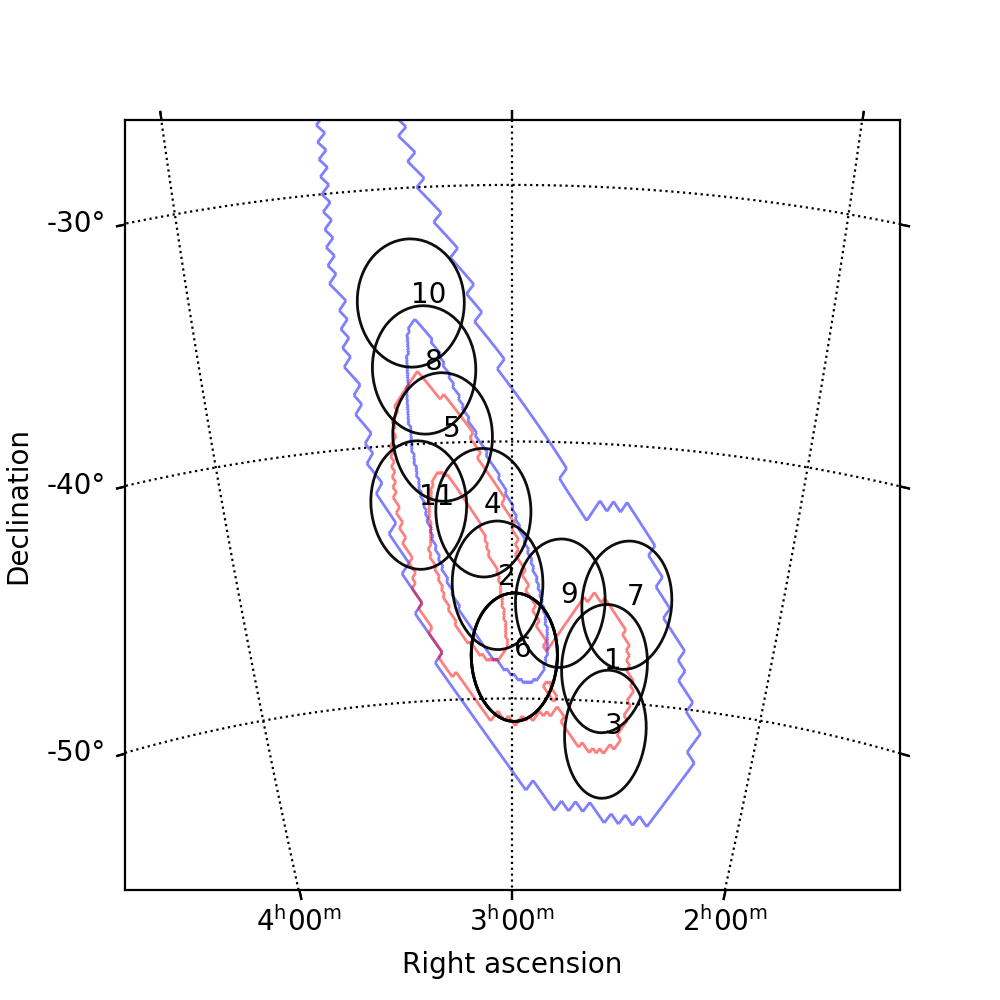}
    \caption{H.E.S.S. coverage of GW170814. The blue contours represent the 90\% and 50\% localization regions for the BAYESTAR maps that triggered the telescopes distributed initially. The red contours represent the 90\% and 50\% localization regions for the updated LALInference maps. The black circles represent the 2.5 degrees FoV of performed observations by chronological order.}
  \label{fig:GW170814_HESS}
  \end{minipage}
  \hfill
  \begin{minipage}[b]{0.44\textwidth}
      \centering
    \includegraphics[width=\textwidth]{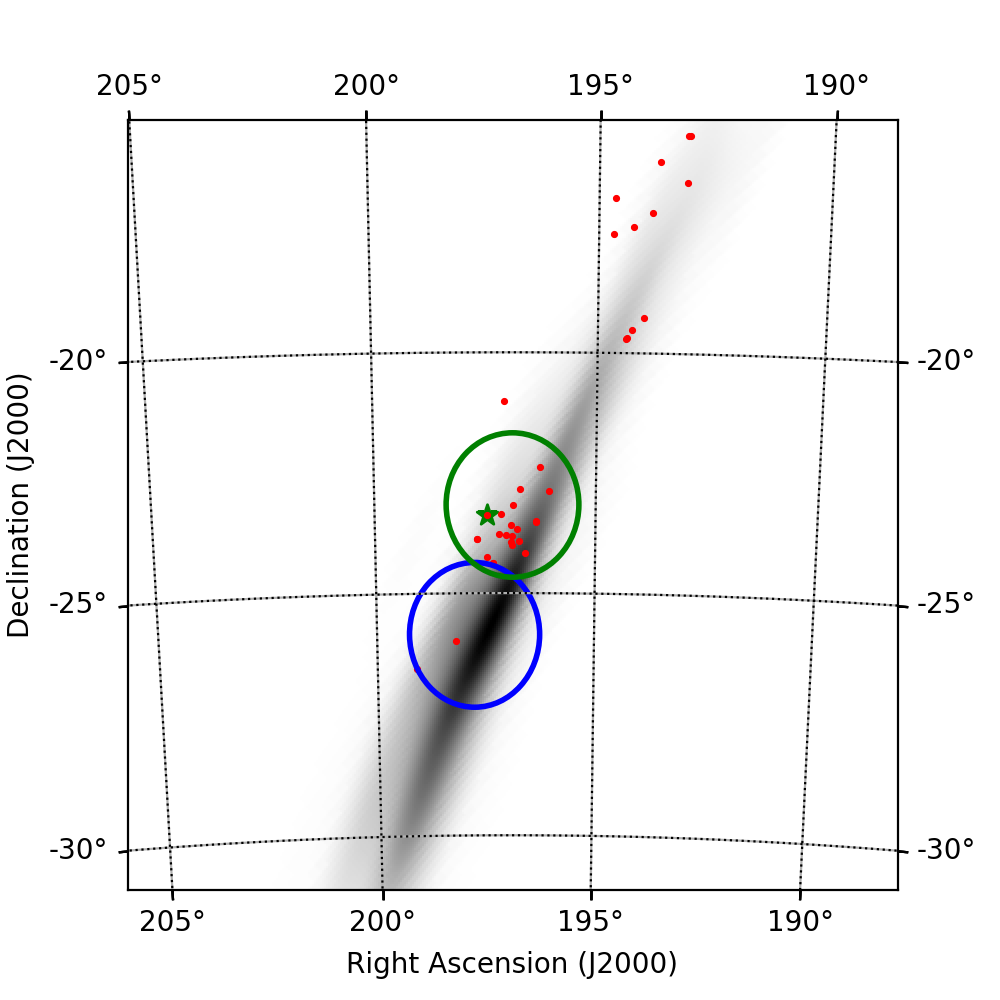}
    \caption{GW170817 first LALInference localisation map. The green and blue circles represents the 1.5 degree radius FoV of the first scheduled observation using 3D algorithms and 2D algorithm respectively. The green star represent the host galaxy NGC4993 and the red dots represent the highest potential host galaxies in the region of the GW event.}
\label{fig:GW1708173D2D}
    \end{minipage}
\end{figure}

GW170814 is a binary black hole merger event that was detected by the two LIGO and the Virgo interferometers~\citep{GW170814}. This was the first time that a GW event was detected by all three observatories and the added independent baselines from Virgo reduced the localization uncertainty significantly. H.E.S.S. started follow-up observations on the $16^{th}$ of August 2017. Observations were obtained during three consecutive nights as a first attempt of science observations with H.E.S.S. following a GW event. The obtained 2D coverage using the \texttt{Best-pixel} algorithm is shown in Fig.~\ref{fig:GW170814_HESS}. Assuming a circular FoV with a radius of 2.5 degree, corresponding to the 12m telescopes, the pointing pattern shown in this figure has been derived. The observations were carried out as planned. The obtained observations cover 95\% of the final localisation map of the event and therefore allowed for the first time to derive meaningful upper limits on the VHE $\gamma$-ray flux. As noted in Sec.~\ref{sec:2Dalgorithms}, the used \texttt{Best-pixel} algorithm is causing significant overlap between the different pointings, a drawback that is corrected by the \texttt{PGW-in-FoV} method. The related improvement is illustrated for example by the fact that a \texttt{PGW-in-FoV} scheduling would have been able to achieve a similar coverage with 2 fewer observations. Preliminary analysis results of the obtained H.E.S.S. data on GW170814 are available at~\citep{GW170814_HESS}. In order to exploit the high sensitivity and low energy threshold, the preliminary analysis of the data is relying on data from the 28m telescope. This leads to a reduction of the effective FoV available for the final results, a fact that has been taken into account subsequently by conservatively assuming a 1.5 degree FoV radius for scheduling GW follow-up observations during O3. Applying this value to GW170814, the VHE coverage of the final localisation area reaches 68\%.\\

Multi-messenger astrophysics with GWs started with the detection of the binary neutron star merger GW170817 on August 17, 2017~\citep{2017PhRvL.119p1101A}. The event identified by LIGO occurred at 12:41:04 UTC. Located at a distance of $40^{+8}_{-14}~\mathrm{Mpc}$, it is still the only GW event that was located at a small enough distance for the efficient use of a galaxy catalog during the scheduling. A BAYESTAR-reconstructed 3D localisation map using data from all three interferometers was published by LVC at 17:54:51 UTC~\citep{GW170817_GCN21513}. The 90\% region of the localization uncertainty had a size of $31~\mathrm{deg}^2$ and was used as input for the \texttt{PGalinFoV} algorithm. H.E.S.S. data taking started on August 17 at 17:59 UTC when the necessary darkness conditions were reached. This was only about 5 minutes after the publication of the localisation of the GW event. Like in Fig.~\ref{fig:GW1708173D2D}, the first pointing position generated by the 3D algorithm covered the host galaxy in spite of it being located at the edge of the skymap away from the GW hotspot. Demonstrating the superiority of the 3D galaxy based approach, a 2D \texttt{PGW-in-FoV} algorithm would naturally focus on the hotspot and would therefore have missed the location of the merger. Thanks to the efficient and rapid reaction, H.E.S.S. was the first ground based facility to get on target and take relevant data of the BNS merger, several hours before the discovery of the optical counterpart. Observations of GW170817 with H.E.S.S. then continued over several days until the direction was not observable any longer. An extensive additional campaign covering the peak of the X-ray emission from the source started 4 months later in December 2017. Analysis and interpretations of these datasets are discussed in~\citep{GW170817_HESS} and~\citep{EM170817_HESS} respectively.\\

\subsection{GW follow-up during O3}
\begin{figure}
  \centering
  \begin{minipage}[b]{0.49\textwidth}
    \includegraphics[width=\textwidth]{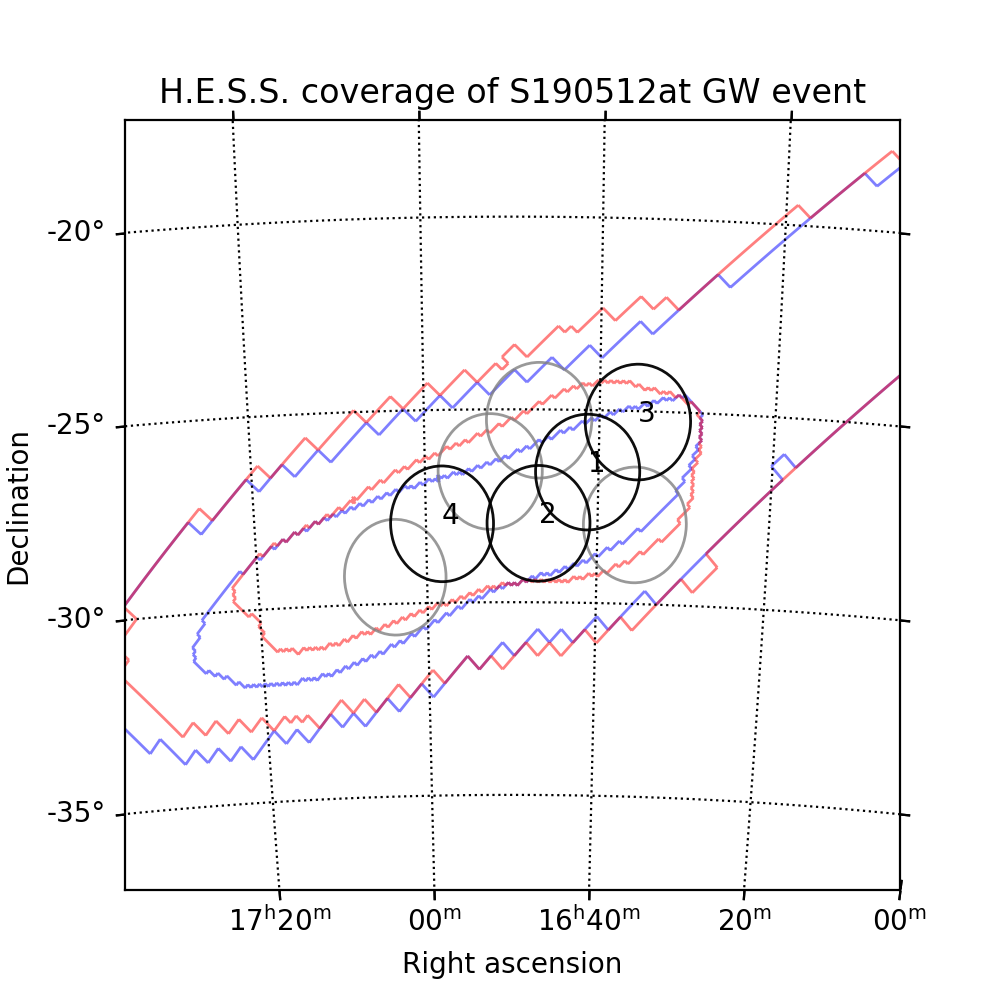}
  \end{minipage}
  \begin{minipage}[b]{0.49\textwidth}
    \includegraphics[width=\textwidth]{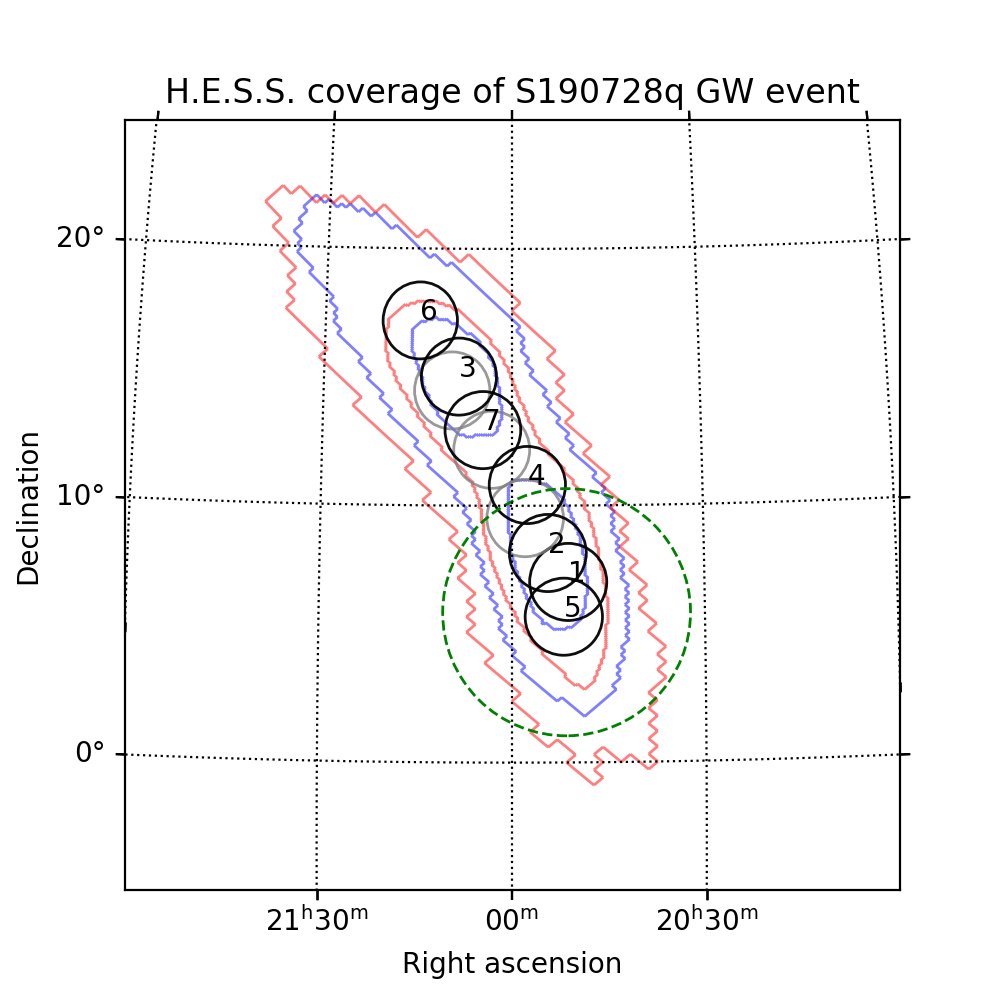}
    \end{minipage}
     \begin{minipage}[b]{0.49\textwidth}
    \includegraphics[width=\textwidth]{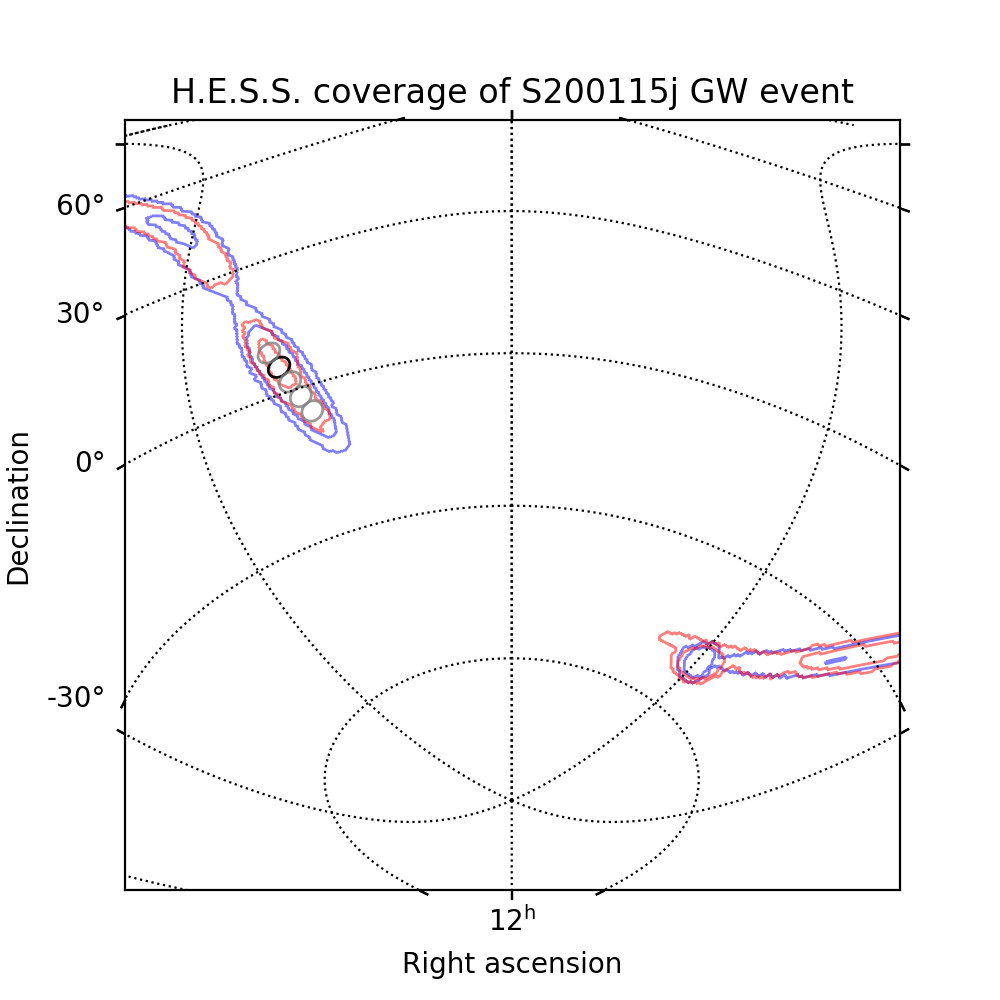}
  \end{minipage}
  \begin{minipage}[b]{0.49\textwidth}
    \includegraphics[width=\textwidth]{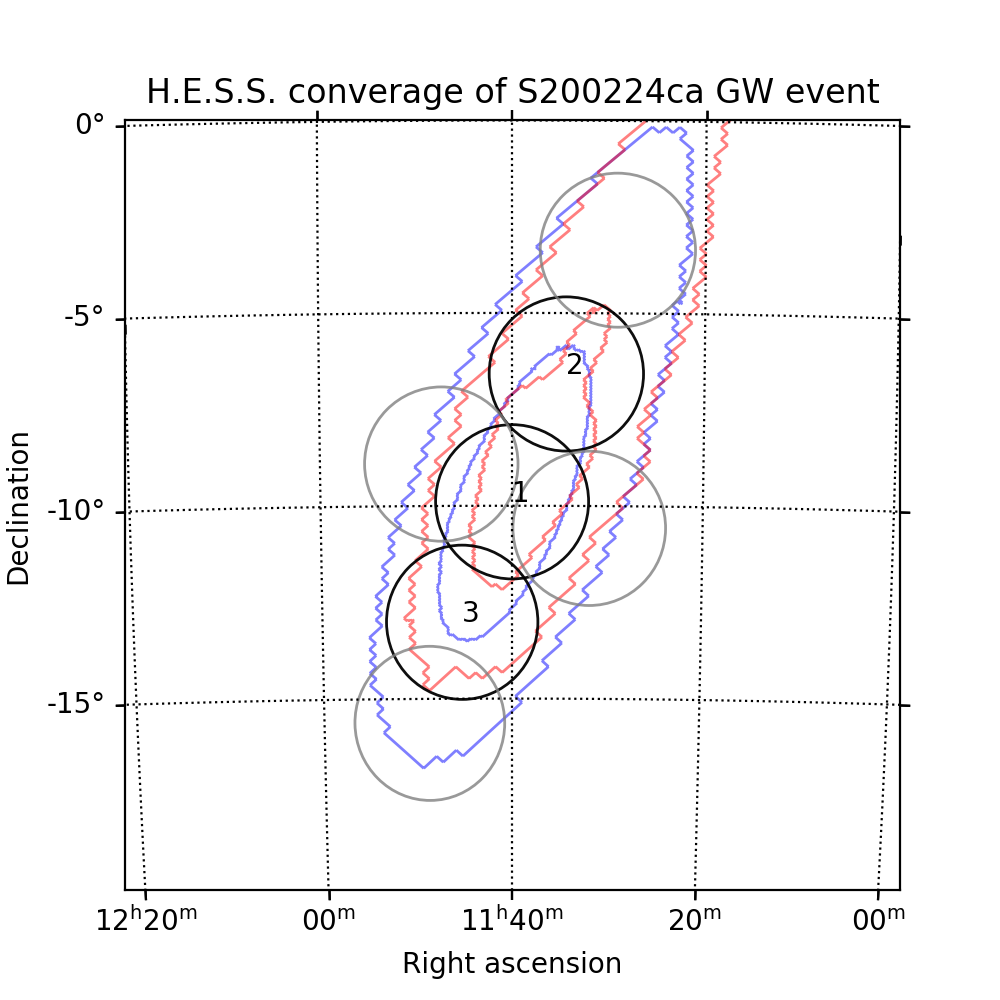}
  \end{minipage}
 \caption{H.E.S.S. coverage of O3 GW events. The blue contours represent the 90\% and 50\% localization regions for the BAYESTAR maps distributed in the initial notices that triggered the telescopes. The red contours represent the 90\% and 50\% localization regions for the updated LALInference maps. The grey circles represent the H.E.S.S. FoV of scheduled observations covering the initially distributed maps and the black circles represent the FoVs of successful observations by chronological order. The FoVs have a radius of 1.5 degrees for S190512at and S190728q and 2 degrees for S200115j and S200224ca. For S190728q the green dashed circle represent the neutrino uncertainty region.}
\label{fig:O3HESScoverage}
\end{figure}

S190512at~\citep{S190512at_initial} is the first GW event detected during the physics run O3 to be well located with a favorable zenith angle for H.E.S.S. Observations scheduled manually in \textit{afterglow} mode covering this event were performed for testing and commissioning purposes. A preliminary version of \texttt{PGW-in-FoV} algorithm has been used to determine the pointing pattern. The planned observations of S190512at covered 34\% ($P_{\text{GW}}$) of the {\it initial} localisation map. Influenced by bad weather not all of the scheduled observations could be obtained, thus leading to a coverage of 21\% of the {\it updated} map~\citep{S190512at_update}. Both, the scheduled and the obtained pointing patterns are illustrated in Fig.~\ref{fig:O3HESScoverage}.\\ 

S190728q is the second event followed by H.E.S.S. during O3. The {\it initial} alert~\citep{S190728q_initial} classifying this event as a NSBH merger was received on 2019-07-28 07:39:04 UTC. Within a minute the H.E.S.S. ToO alert system, evaluated the possibility of a \textit{prompt} reaction and sent out a notice to the H.E.S.S. GW team. Four observation runs were scheduled to be obtained later that night covering ~32\% of the {\it initial} skymap. At 2019-07-28 20:29:12 UTC an {\it update}~\citep{S190728q_update} was received after the first follow-up observation had already started and a new schedule containing 6 new observational positions was automatically distributed. A total of seven observation runs were obtained. These covered a part of the uncertainty region of a nearby neutrino alert that was emitted by the IceCube collaboration during the same day~\citep{S190728q_neutrino}. Initially, the experts on call have foreseen to schedule four additional runs on the optical transient (OT) ZTF19abjethn~\citep{S190728q_ztf}. They were cancelled after the dissociation~\citep{S190728q_ztf_notassociated} of the OT to the GW event. All scheduled positions were observed by H.E.S.S.~\citep{S190728q_HESS} and allowed to cover 64\% of the total $P_{\text{GW}}$ of the {\it updated} GW localisation region. S190512at and S190728q are confirmed as BBH merger events and their properties can be found in the GWTC-2 catalog~\citep{GWTC-2} under GW190512\_180714 and GW190728\_064510 respectively.\\

In the end of 2019 the large 28m H.E.S.S. telescope underwent a camera upgrade and observations were carried out by the remaining four 12m small telescopes that have a larger FoV. Taking this into consideration, the FoV parameter in the H.E.S.S. GW follow-up configuration has been changed from 1.5 to a conservative 2 degrees radius for the remainder of O3. Moreover, following re-assessement by the observations committee, the minimum allowed $P_{\text{GW}}$ coverage per observation has been reduced to 2\% (cf. Sec.~\ref{sec:program}) for mergers involving a NS. With these new conditions H.E.S.S. managed to observe two additional GW events before the early stop of O3 on March 27, 2020.\\

S200115j is a NSBH merger detected on 2020-01-15 at 04:23:09 UTC~\citep{S200115j_initial}. Five observations were derived by the \texttt{PGW-in-FoV} algorithm. These would have allowed to cover 25\% of the {\it initial} localisation region. Due to bad weather, only one run could be taken. It covers 2.4\% of the {\it updated} map~\citep{S200115j_update}.\\

The last GW event observed by H.E.S.S. before the end of O3 is S200224ca~\citep{S200224ca_initial}, a BBH merger. The alert arrived during the night but because of bad weather the telescopes were parked in and \textit{prompt} observations were not possible. The follow-up started $\sim$3 hours later, thus cutting short the schedule that had foreseen seven observations. Nevertheless, the three successful runs cover 72\% of the {\it initial} and 70\% of the {\it updated} localisation map~\citep{S200224ca_update}, making S200224ca the BBH merger with the highest coverage in the VHE domain.\\



\section{Discussion and conclusion}
\label{sec:discussion}
The H.E.S.S. GW follow-up program was successfully implemented and tested during O2 and O3 and it will continue to allow automatic follow-up in the VHE $\gamma$-ray domain of the most promising GW events accessible from the southern hemisphere in the upcoming observing runs. The implemented 3D strategies for deriving the follow-up observation schedules relying on targeting local galaxies have proven their efficiency in the successful coverage of the GW170817 electromagnetic counterpart before its discovery in optical observations. 2D strategies are better adapted for the scheduling of the remaining 5 GW events observed by H.E.S.S. during O2 and O3. This is driven by the large distance of these GW events and the lack of completeness of currently available galaxy catalogs. GW170817 remains the only GW events with an identified EM counterpart. H.E.S.S. coverage of the observed events is higher than 50\% in most cases and reaches 70\% in the case of S200224ca assuming a conservative FoV radius. While the follow-up schedule has been derived based on the low latency and preliminary GW localisations, the pointing pattern used for the H.E.S.S. observations combined with a relatively large FoV are robust against changes in the GW event reconstructions and thus allowing to cover significant portions of the final, offline reconstructed GW localisation uncertainty regions. The expected H.E.S.S. sensitivity from one standard observation run is of the order of a few $10^{-12}$ erg cm$^{-2}$ s$^{-1}$~\citep{GW170817_HESS} in the VHE domain ranging from 100s of GeVs to TeV energies. Analysis results will be published in a forthcoming publication~\citep{hessResultsPaper}. 

We have also optimised the capabilities of the GW follow-up plugin within the H.E.S.S. ToO alert system for reaction speed and will thus be able to fully benefit from ongoing efforts by LVC to provide automated alerts with even shorter latency than the {\it preliminary} alerts emitted during O3. For exceptional CBC events with a high signal-to-noise ratio, it may be possible for GW detectors to detect the event already during the inspiral phase before the merger itself and issue a pre-merger alert~\citep{earlyWarning}. These \textit{early warnings} are particularly beneficial in the search for GRB-like counterparts from BNS mergers in VHE domain with IACTs as observations during the prompt or early afterglow phase promise a rich dataset possibly shedding light on the central engine driving the VHE emission. The H.E.S.S. estimated prompt reaction time to the most promising GW events was less than 1 minute for the O3 period with a total maximum latency of 2 minutes between the reception of the alert and the beginning of data taking. Further time saving measures are currently being implemented. They include loading the galaxy catalog at the beginning of each night to save the time required by that step and general optimisation and parallelisation of the scheduling framework. On the other hand, recent detections of VHE emission from (long) GRBs by both the H.E.S.S.~\citep{GRB180720B, GRB190829A_HESS} and MAGIC~\citep{GRB190114C} IACTs provide significant insights in the duration of the VHE emission of these events. Assuming that the flux decay in the afterglow of short GRBs caused by BNS mergers follows a similar behaviour, we may be able to detect the associated VHE emission over several hours or even days with current and future IACTs. The H.E.S.S. follow-up strategy of GW events will be adapted accordingly in preparation of the upcoming observation run O4. 

The lack of GW counterpart detection during O3 hints to the scarcity of events like GW170817. Continuous optimisation of counterpart search strategies beyond the speed improvements and extension of the time coverage of the observations is therefore crucial for effective hunting for electromagnetic emission from such \textit{golden} events. Improvements to the H.E.S.S. GW follow-up program include the possible use of galaxy stellar masses provided in catalogs like MANGROVE~\citep{MANGROVE}. This option was already available in the offline tools used by the H.E.S.S. GW experts team during O3 and is now being implemented in the automatic follow-up scheme. The MANGROVE approach uses a weighing parameter to prioritize observations of massive galaxies over less massive ones in the scheduling. 

Another extension of the H.E.S.S. framework is related to un-modelled burst events. While the burst pipelines~\citep{2011PhRvD..83j2001K} are also sensitive to CBC events, the dedicated template based CBC pipelines typically provide better sensitivities and higher signal-to-noise ratios to these events. Out of the many events that could be caught primarily by the burst searches, nearby, i.e. Galactic, supernovae are one of the most promising ones~\citep{2016PhRvD..93d2002G}. The pipelines searching for un-modelled signals provide only a 2-dimensional localisation. Taking into account the limited horizon of these searches, we foresee to correlate the GW localization uncertainty of {\it burst} alerts with the Galactic plane. Moreover, the general H.E.S.S. system for ToOs and follow-up observations of internal and external alerts will be improved by adding effective methods searching for transient signals in the real-time data stream. Examples of such methods are described in~\citep{Brun_transient_tools}. 

Finally, the flexibility of the methods described in this paper allows them to be applied to any IACT like the future Cherenkov Telescope Array (CTA) by adopting the right high level telescope parameters like observatory location and FoV. In fact, the advanced capabilities of CTA with a large FoV, a low energy threshold, a higher sensitivity than current IACTs, and a large number of telescopes potentially permitting operations in several sub-arrays will allow it to be a very efficient GW follow-up machine. Furthermore, the Japanese KAGRA and an additional LIGO interferometer in India are currently under commissioning and construction respectively and are planned to join the search for GWs effort in the future. The addition of two GW detectors is expected to further improve GW localisations and increase the chances of counterpart detection.  

\section*{Acknowledgement}
We would like to thank the members of the H.E.S.S. Collaboration for their technical support and for helpful discussions. The results and algorithms described in this paper have been derived using Python~\citep{python} primarly with the HEALPix/healpy~\citep{healpy1}~\citep{healpy2}, Astropy~\citep{astropy:2018}, Numpy~\citep{numpy}~\citep{numpy_nature}, Matplotlib~\citep{matplotlib}, PyEphem~\citep{pyephem}, and MOCPy~\citep{mocpy} packages.


\bibliographystyle{JHEP}
\bibliography{mybibfile}


\end{document}